\documentclass{jfm}

\usepackage{color}
\usepackage{amsmath}
\usepackage{amsfonts}
\usepackage{amssymb}

\usepackage{graphicx}
\usepackage{natbib}
\usepackage{tikz,pgfplots}
\usepackage{epstopdf, epsfig}

\newcommand{\qb}{ \boldsymbol{q} }
\newcommand{\qtb}{ \tilde{\boldsymbol{q}} }
\newcommand{\Ab}{ \boldsymbol{A} }
\newcommand{\Lb}{ \boldsymbol{L} }

\newcommand{\qhb}{ \hat{\boldsymbol{q}} }

\newcommand\ci{\text{i}}            
\newcommand\cc{\text{c.c.}}            

\shorttitle{DNS and modal analysis of flow over swept airfoil sections}
\shortauthor{N. De Tullio and N. D. Sandham}

\title{Direct numerical simulations and modal analysis of subsonic flow over swept airfoil sections}

\author{Nicola  De Tullio\aff{1}
   and Neil D. Sandham\aff{1} \corresp{\email{n.sandham@soton.ac.uk}} }

\affiliation{\aff{1}Aerodynamics and Flight Mechanics Group, University of Southampton, Southampton SO17 1BJ, UK}

\begin{document}

\maketitle

\begin{abstract}
Direct numerical simulations (DNS) and modal analysis techniques are applied to investigate the flow over a NACA-0012 airfoil at $\Rey = 50,000$. Three different sweep angles are considered, namely $\Lambda=0^\circ$, 20$^\circ$ and 40$^\circ$, for two sweep configurations. Using models for the separation bubbles, Reynolds number and thickness effects are separated from sweep effects. The transitional flow structure is observed to change with sweep angle, with swept cases showing more spanwise-coherent large structures. At $\Lambda=20^\circ$ these structures are perpendicular to the free stream direction, whereas at $\Lambda=40^\circ$ they are parallel to the leading edge. A good agreement between Fourier analysis of the DNS data and global stability analysis suggests that the changes are due to the emergence in the $\Lambda=40^\circ$ swept case of an unstable global mode. The global mode has coupled acoustic and vortical support, implying a coupling between trailing edge sound production and shear layer instability. Dynamic mode analysis shows the presence of lower frequency non-acoustic modes in the highly swept case that are not present in the unswept case.

\end{abstract}


\section{Introduction}
Transitional separation bubbles (TSB) occur when laminar boundary-layer flow over a solid surface separates under the influence of a sufficiently high adverse pressure gradient, generating a region of reversed flow and a highly unstable detached shear layer. The shear layer develops instabilities that drive the flow to a chaotic state and eventually cause the flow to reattach as a turbulent boundary layer. TSBs are typical of moderate Reynolds number flows over airfoils, where they exert a strong influence on aerodynamic performance. The pocket of ``dead air" that forms inside the bubble limits the maximum lift attainable, while the characteristics of the separation bubble as a whole determine the stall behaviour of the airfoil. Numerous research efforts have been carried out to better understand the behaviour of TSBs since the first reported observations of \citet{jones34}. Early experimental investigations focused mainly on the phenomenon of bubble bursting \citep[][]{gault49,mccullough51,gaster67,horton69}, whereby the detached shear layer placed over the top of a short separation bubble may suddenly fail to reattach, creating a long separation bubble and leading to airfoil stall. These early investigations, together with more recent ones \citep[][]{pauley90,diwan06,sandham08,marxen11}, have led to a number of possible physical explanations for bubble bursting and semi-empirical correlations that are able to collapse the available experimental data. 

The structure and dynamics of TSBs, including their bursting behaviour, are greatly influenced by the underlying laminar-turbulent transition process that shapes them. A number of experiments \citep[][]{dovgal94,watmuff99} and numerical simulations \citep[][]{rist02,marxen12,marxen13} have highlighted the importance of the Kelvin-Helmholtz (K-H) instability of the separated shear layer as a dominant process in TSBs. These investigations place an emphasis on the ability of separation bubbles to strongly amplify incoming disturbances. However, self-sustained flow unsteadiness and laminar-turbulent transition in separation bubbles have also been observed in various numerical investigations \citep[see for example][]{pauley90,marquillie03,postl11}, indicating that the portrayal of TSBs as disturbance amplifiers does not always suffice. The resonator character of TSBs has been investigated in the context of absolute and global stability analyses \citep[][]{huerre90} by a number of authors. Investigations of absolute instability in separation bubbles \citep[][]{hammond98,alam_sandham, rist02} suggest that a reversed flow magnitude of at least $15-30\%$ of the free stream is needed before an absolutely unstable mode associated with the K-H instability of the separated shear layer appears. Such a mode appears to lead to self-sustained two-dimensional vortex shedding over which an additional self-excited three-dimensional (3D) instability may evolve \citep[][]{jones_sandham08,embacher14}, driving the flow to a turbulent state. This additional 3D mode may be viewed as an unstable global Floquet mode of the periodic flow induced by the saturated K-H instability.

The global stability of nominally two-dimensional (2D) separation bubbles has been investigated for a variety of different flow configurations, including flat plate boundary layers \citep[][]{theofilis00,rodriguez10,rodriguez13}, backward facing steps \citep[][]{barkley02}, curved channels \citep[][]{marquet08,marquet09}, roughness elements \citep[][]{gallaire07,ehrenstein08} and shock/boundary-layer interactions \citep[][]{robin07}. In all these cases it was found that, in the absence of external disturbances, the primary instability of the separation bubble is a steady three-dimensional global centrifugal mode \citep[first found by][]{theofilis00} that becomes unstable for a reverse flow of about $7\%$; well below the values reported for the onset of the unsteady, absolutely unstable K-H mode. The work of \citet{ehrenstein08} shows that, in addition to the steady three-dimensional mode, separation bubbles induced by two-dimensional roughness elements can also sustain the growth of two-dimensional globally unstable modes, the non-normality of which may lead to low frequency self-sustained flapping.

Starting with Theofilis et al (2002), techniques to study the global stability of airfoil flows have been developed, with low-Reynolds-number examples for stalled flow over airfoils in \citet{kitsios09} and \citet{zhang16}. In particular, \citet{rodriguez11} associated the formation of three-dimensional stall cells with a stationary global mode.  Other applications of the technique have been in two main contexts: transonic buffet and airfoil tonal noise. The global stability of the compressible transonic flow over a NACA0012 airfoil at transonic-buffet conditions was studied by \citet{crouch02}. The linearised dynamics of the flow were analysed with respect to steady solutions of the Reynolds Averaged Navier-Stokes (RANS) equations, and included a linearised turbulence model for the averaged Reynolds stresses. The results obtained using global stability analysis for the onset of buffet were found to be in good agreement with experimental results. \citet{sartor15} also used RANS and global stability analysis to investigate the transonic flow over the OAT15A supercritical airfoil, showing that the buffet phenomenon is the consequence of a global instability of the flow. Their results also shows that the shock/separation bubble interaction behaves as a low pass filter, damping out high frequency disturbances. At lower Mach numbers and Reynolds numbers, airfoils can exhibit a phenomenon of tonal noise (\citet{nash99}, \citet{plogmann13} and \citet{probsting15}) that is explained as a feedback loop containing trailing edge noise radiation coupled to convective shear layer instabilities (see for example \citet{arbey83}, \citet{desq07} and \citet{chong12}). \citet{jones11} showed that the flow around the airfoil can be rendered globally unstable (or stable, depending on Re, M and angle of attack) by an acoustic feedback loop sustained by the scattering of acoustic waves generated by the interaction between convectively unstable modes growing over the bubble and the airfoil's trailing edge. More recently, the instability of the subsonic flow over a NACA0012 airfoil was investigated by \citet{fosasdepando14}, showing that tonal noise in airfoils is the manifestation of a branch of global modes characterised by an acoustic feedback loop originating at the airfoil's trailing edge. \citet{fosas17} studied adjoint global modes, which were found to have compact support in the boundary layer upstream of the separated flow region. The overlap between the direct and adjoint modes was then associated with a wavemaker region that could be an optimal location for actuators to influence the tonal noise characteristics.

Most of the research on TSBs in the past few decades has been directed towards the unswept configuration, while the effects of sweep on the TSB behaviour have received little attention. Experimental investigations \citep{young66, hortonthesis} concentrated on relating the mean flow properties of swept separation bubbles with the corresponding unswept configuration, thereby exploring the applicability of the so called \emph{independence principle}, according to which the introduction of infinite sweep does not affect the flow characteristics in the chordwise direction. On this basis, \citet{davis87} were able to extend the semi empirical method to infinite swept wings. More recently, \citet{kaltenbach00} performed direct numerical simulations of transitional swept separation bubbles behind a rearward-facing step and showed that the independence principle holds throughout the entire separated flow region for sweep angles up to $40^\circ$. However, the introduction of disturbances to the otherwise steady laminar inflow had the effect of decreasing the range of sweep angles for which the independence principle is applicable. Instability growth rates (extracted directly from the DNS data) in the detached shear layer induced by the bubble where found to grow slightly with sweep angle. The independence principle was shown to hold exactly in the case of strictly laminar swept separation bubbles by \citet{hetsch09a}, provided the free-stream velocity in the chordwise direction remains independent of sweep angle. In a follow-up study, \citet{hetsch09b} also investigated the linear stability of laminar swept separation bubbles induced over a flat plate by an adverse pressure gradient. Using local linear stability analysis (LST) and solutions of the parabolised stability equation (PSE) they showed that the linear stability results for swept separation bubbles are not independent of sweep angle. In particular, the primary instability of the separation bubble was found to shift towards higher frequencies and spanwise wavenumbers with increasing sweep angle, while the dominant Tollmien-Schlichting wave in the attached boundary layer was only slightly affected by sweep. Despite cross-flow levels of up to $9\%$ of the free-stream velocity, \citet{hetsch09b} found that cross-flow instabilities played only a marginal role in the laminar-turbulent transition process, and only for the highest sweep angle of $\Lambda = 45^\circ$ analysed in their study.

In this work we study how sweep affects the laminar-turbulent transition of the separation bubbles developing over a NACA-0012 airfoil in the compressible subsonic regime at moderate Reynolds numbers. The investigation is carried out using DNS and global linear stability analysis. The remainder of the paper is organised as follows. In \S \ref{sec:math_and_numerics} we present the mathematical and numerical details of the DNS and global stability calculations, while the simulation parameters and flow configurations analysed are explained in detail in \S \ref{sec:flow_configuration}. A discussion of the DNS results is provided in \S \ref{sec:mean_flow} for the mean flow features and \S \ref{sec:unsteady_features} for the unsteady flow characteristics. An analysis of the dynamic mode decomposition of the DNS results is provided in \S \ref{sec:dmd_global}, along with the results of a two-dimensional global stability analysis of the mean airfoil flows. The paper ends in \S \ref{sec:conclusions} with the main conclusions of the investigation.
 
\section{Mathematical and numerical details}\label{sec:math_and_numerics}
\subsection{Direct numerical simulations}
Direct numerical simulations are carried out using the SBLI code; a high-order, multi-block, finite-difference solver developed at the University of Southampton. The code solves both the nonlinear and the linearised compressible Navier-Stokes equations. In conservative, dimensionless form the nonlinear Navier-Stokes equations are written as
\begin{subequations}
\label{eq:ns}
\begin{align}
\label{eq:ns_1}
&\frac{\partial \rho}{\partial t}+\frac{\partial \rho u_j}{\partial x_j}=0, \\
\label{eq:ns_2}
&\frac{\partial \rho u_i}{\partial t} +\frac{\partial \rho u_i u_j}{\partial x_j}+\frac{\partial p}{\partial x_i}=\frac{\partial \tau_{ij}}{\partial x_j},  \\ 
\label{eq:ns_3}
&\frac{\partial \rho E}{\partial t} +\frac{\partial \left(\rho E +p\right)u_i}{\partial x_i}=-\frac{\partial q_i}{\partial x_i}+\frac{\partial u_i\tau_{ij}}{\partial x_j},
\end{align}
\end{subequations}
where the symmetric viscous stress tensor, $\tau_{ij}$, is defined as
\begin{equation}
\tau_{ij}=\frac{\mu}{\Rey}\left(\frac{\partial u_j}{\partial x_i}+\frac{\partial u_i}{\partial x_j}-\frac{2}{3}\frac{\partial u_k}{\partial x_k}\delta_{ij}\right),
\end{equation}
and $\delta_{ij}$ is the Kronecker delta function defined as $\delta_{ij}=1$ for $i=j$ and $\delta_{ij}=0$ for $i\neq j$. 
The above equations are closed using the equation of state for the calculation of the pressure and Fourier's law of heat conduction for the calculation of the heat flux vector, given respectively by
\begin{equation}
p=(\gamma-1)\left(\rho E-\frac{1}{2} \rho u_iu_i\right)=\frac{1}{\gamma M^2}\rho T  \hspace{0.3cm} \text{and} \hspace{0.3cm}q_j=-\frac{\mu}{(\gamma-1)M^2\Pran \Rey}\frac{\partial T}{\partial x_j}. 
\end{equation}
Here, the dynamic viscosity is calculated from the temperature field using the Sutherland's law $\mu=T^{3/2}(1+S^*/T_r^*)/(T+S^*/T_r^*)$, where $S^*=110.4$ K is the Sutherland constant for air and $T_\infty^*=273.15$ K. Dimensionless variables are obtained as
\begin{align}
&t=\frac{t^* U_\infty}{c^*}, \hspace{2mm} x_i=\frac{x_i^*}{c^*}, \hspace{2mm} u_i = \frac{u_i^*}{Q^*_\infty}, \hspace{2mm} \rho = \frac{\rho^*}{\rho^*_\infty}, \\
& p = \frac{p^*}{\rho^*_\infty Q^{*2}_\infty}, \hspace{2mm} \mu = \frac{\mu^*}{\mu^*_\infty }, \hspace{2mm} E = \frac{E^*}{Q^{*2}_\infty },
\end{align}
where asterisks (*) denote dimensional quantities, while $c^*$ and $Q^*_\infty$ are the dimensional airfoil chord and free-stream velocity, respectively. Here, $t^*$ is time, $(x^*_1, x^*_2, x^*_3) = (x^*, y^*, z^*)$ are the streamwise (perpendicular to the leading edge of the airfoil), wall-normal and spanwise components of the position vector, $(u^*_1, u^*_2, u^*_3) = (u^*, v^*, w^*)$ are the streamwise (perpendicular to the leading edge of the airfoil), wall-normal and spanwise components of the velocity vector, $\rho^*$ is the density, $p^*$ is the pressure, $\mu^*$ is the dynamic viscosity and $E^*$ is the total energy per unit mass. The dimensionless parameters which define the problem are the Reynolds number $\Rey = \rho^*_\infty Q^*_\infty c^*/\mu^*_\infty$, the Mach number \textit{M} and the Prandtl number $\Pran=0.72$. 

The SBLI code uses a standard fourth-order central difference scheme to calculate derivatives at internal points and a stable treatment developed \citet{carp} for the calculation of derivatives at domain boundaries. Time integration is based on a third-order compact Runge-Kutta method \citep{rk}. In its nonlinear variant, the code employs an entropy splitting approach, whereby the inviscid flux derivatives are split into conservative and non-conservative parts. The entropy splitting scheme, together with a Laplacian formulation of the heat transfer and viscous dissipation terms in the momentum (\ref{eq:ns_2}) and energy (\ref{eq:ns_3}) equations (which prevents the odd-even decoupling typical of central differences), helps improve the stability of the low dissipative spatial discretisation scheme used. More details of the basic scheme are given in \citet[][]{entr}. Code parallelisation is achieved using the Message Passing Interface (MPI) library.  

\subsection{Global stability analysis}
An investigation of the importance of linear dynamics in the laminar-turbulent transition process is carried out through global linear stability analysis of the time- and span-averaged flows. To this end, here we use the SBLI code in a linearised mode, which solves the linearised compressible Navier-Stokes equations written as
\begin{subequations}
\label{eq:nsLin}
\begin{align}
\label{eq:nsLin_1}
&\frac{\partial \rho'}{\partial t}+\frac{\partial \bar{\rho} u'_i}{\partial x_i}+\frac{\partial \rho' \bar{u}_i}{\partial x_i}=0 \\
 \label{eq:nsLin_2}
&\frac{\partial u'_i}{\partial t} + \left( \frac{\rho'}{\bar{\rho}}\bar{u}_j + u'_j \right)\frac{\partial \bar{u}_i}{\partial x_j} + \bar{u}_j \frac{\partial u'_i}{\partial x_j} + \frac{1}{\bar{\rho}}\frac{\partial p'}{\partial x_i} =\frac{1}{\bar{\rho}}\frac{\partial \tau'_{ij}}{\partial x_j}  \\ 
\label{eq:nsLin_3}
&\frac{\partial T'}{\partial t} + \bar{u}_i \frac{\partial T'}{\partial x_i} +\left( \frac{\rho'}{\bar{\rho}}\bar{u}_i+u'_i \right)\frac{\partial \bar{T}}{\partial x_i} + \mathcal{B} \left( \bar{p}\frac{\partial u'_i}{\partial x_i} + p'\frac{\partial \bar{u}_i}{\partial x_i} \right)=-\mathcal{B} \frac{\partial q'_i}{\partial x_i}+ \mathcal{B} \mathcal{D}',
\end{align}
\end{subequations}
where $\mathcal{B} = \gamma(\gamma-1)M^2/\bar{\rho}$. The unsteady pressure $p'$ and the linearised heat fluxes $q'_i$ are written as
\begin{align}
p' = \frac{1}{\gamma M^2}\left( \bar{\rho} T' + \rho' \bar{T} \right) \hspace{0.3cm} \text{and} \hspace{0.3cm} q'_i =-\frac{1}{(\gamma-1)M^2\Pran \Rey}\left( \bar{\mu}\frac{\partial T'}{\partial x_i} + \mu' \frac{\partial \bar{T}}{\partial x_i}\right),
\end{align}
while the components $\tau'_{ij}$ of the linear viscous stress tensor and the linear viscous dissipation term $\mathcal{D}'$ are given respectively by
\begin{align}
\tau'_{ij} = \frac{1}{\Rey}\left \{ \bar{\mu} \left( \frac{\partial u'_i}{\partial x_j} + \frac{\partial u'_j}{\partial x_i} -\frac{2}{3}\frac{\partial u'_k}{\partial x_k}\delta_{ij} \right) + \mu' \left( \frac{\partial \bar{u}_i}{\partial x_j} + \frac{\partial \bar{u}_j}{\partial x_i} -\frac{2}{3}\frac{\partial \bar{u}_k}{\partial x_k}\delta_{ij} \right) \right \}
\end{align}
and 
\begin{align}
\mathcal{D}' = \frac{\partial \bar{u}_i}{\partial x_j} \tau'_{ij} + \frac{\partial u'_i}{\partial x_j} \bar{\tau}_{ij}.
\end{align} 
The linearised SBLI code uses the same spatial and temporal discretisation schemes of its nonlinear variant, as well as the MPI parallelisation routines. The code also retains the Laplacian formulation for the heat transfer term in the momentum (\ref{eq:nsLin_2}) and energy (\ref{eq:nsLin_3}) equations and for the viscous dissipation term in the momentum equation. Both the nonlinear and linearised SBLI solvers have been extensively validated \citep[see for example][]{nicothesis,nicosandham15,nicoruban15}.

The linearised compressible Navier-Stokes equations (\ref{eq:nsLin}) can be written as an initial-value problem (note that, from now on, repeated indices do not entail summation)
\begin{subequations}
\label{eq:linear_ivp}
\begin{align}
\label{eq:linear_ivp_1}
&\frac{\partial \qtb}{\partial t} = \mathcal{A}(\qb_b;k_z)\qtb, \\
\label{eq:linear_ivp_2}
&\qtb = \qtb_0 \hspace{2mm} \text{at} \hspace{2mm} t = t_0,
\end{align}
\end{subequations}
where, making use of the flow homogeneity in the spanwise direction $z$, we have put $\qb(x,y,z,t) = \qtb(x,y,t)\exp(\ci k_z z) + \cc$,
with $k_z \in \mathbb{R}$. Here, $\mathcal{A}$ is the linearised Navier-Stokes operator written explicitly in (\ref{eq:nsLin}) and $\qb = [\rho', u', v', w', T']$ is the linearised solution vector. By writing the linear disturbances in normal mode form $\qtb(x,y,t) = \qhb(x,y)\exp(\omega t)$, where $\omega$ is the complex frequency, and discretising (\ref{eq:linear_ivp}) one arrives at the eigenvalue problem 
\begin{align}
\label{eq:eigvp}
\omega_j \qhb_j = \Ab \qhb_j,
\end{align}
where $\Ab$ is the (complex for $k_z \neq 0$) linearised compressible Navier-Stokes operator matrix and $\omega_j$ and $\qhb_j$ are the complex eigenvalues and corresponding eigenvectors, respectively. In this work, the global eigenvalue problem is solved in matrix-free mode \citep[see for example][]{eriksson_1985, barkley02, bagheri09}, using the Implicitly Restarted Arnoldi Method provided by the parallel ARPACK library \citep{arpack}, with SBLI's linearised compressible Navier-Stokes solver as a timestepper. The timestepper advances the solution vector from a time $t = t_0$ to a time $t=t_0+\tau$ by numerically integrating (\ref{eq:linear_ivp}), leading to the solution
\begin{align}
\qtb(t_0+\tau) = \Lb\qtb_0,
\end{align}
where the evolution operator $\Lb$ is a numerical approximation of the matrix exponential $\displaystyle\exp \left[\Ab\tau \right]$. Therefore, the time stepping approach used in this work leads to the eigenvalue problem 
\begin{align}
\lambda_j \qhb_j = \Lb \qhb_j,
\end{align}
with eigenvalues $\lambda_j = \exp \left(\omega_j \tau \right)$. Here, $\tau$ is the sampling period used in the timestepping Arnoldi procedure and is chosen according to the frequency of the most unstable global modes; its value should be small enough to avoid aliasing (Nyquist sampling theorem) while large enough for the Arnoldi procedure to converge. Note that using the timestepper approach to solve the global eigenvalue problem is equivalent to applying an exponential transformation to (\ref{eq:eigvp}). This transformation maps highly stable eigenvalues $\omega_j$ to $|\lambda_j| \to 0$ and highly unstable ones to $|\lambda_j| \to \infty$, and hence the most unstable global modes are obtained by searching for the $\lambda_j$ with the largest magnitude. 

\section{Flow configurations and simulation parameters}\label{sec:flow_configuration}
\begin{figure}
  \centerline{\includegraphics[trim = 0mm 0mm 0mm 0mm,clip,width=0.9\textwidth]{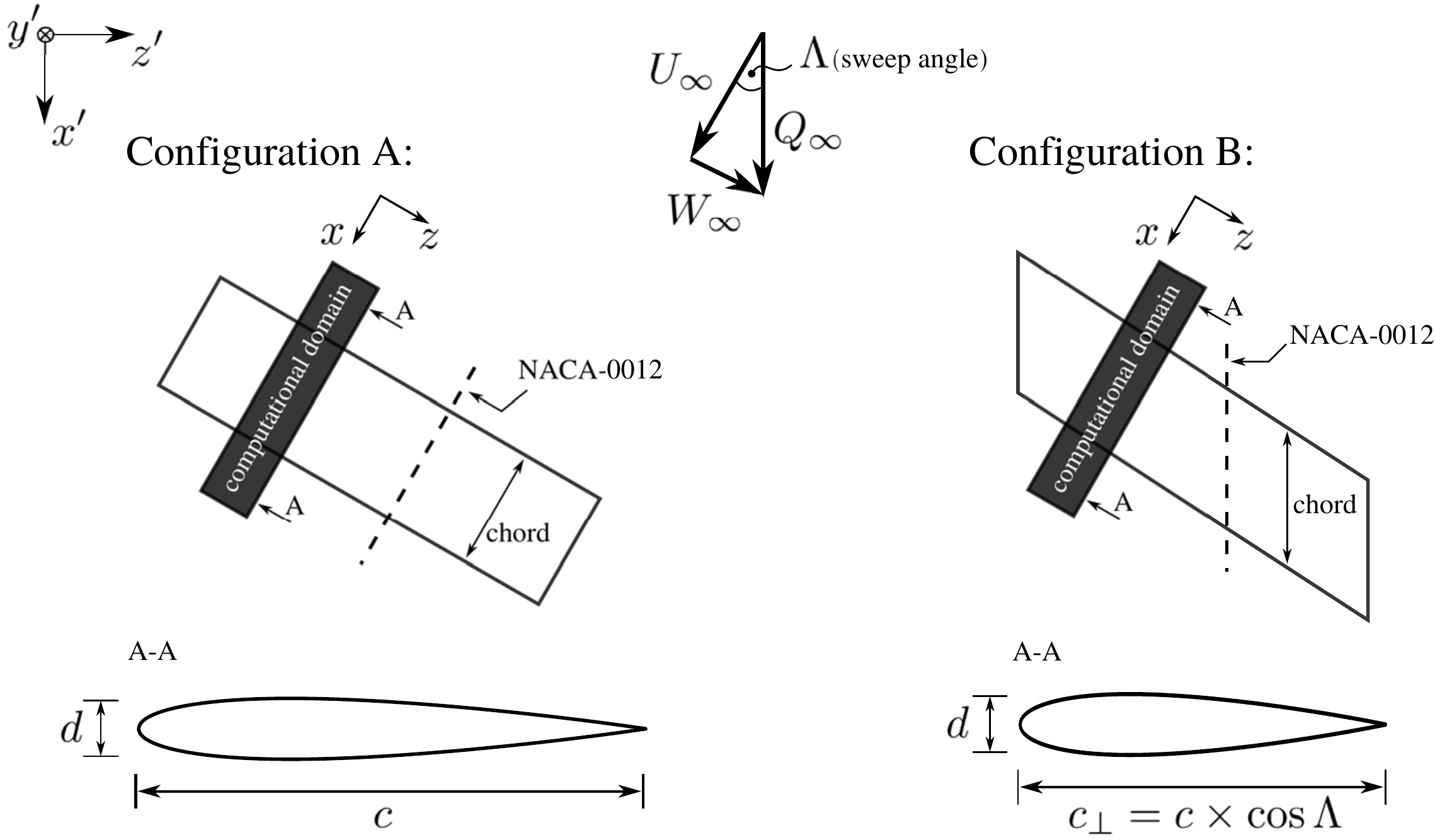}}
  \caption{A sketch of the swept-wing configurations considered.}
\label{fig:config_sketch}
\end{figure}

Direct numerical simulations have been performed for the flow over an infinitely swept wing with a NACA-0012 profile at an incidence $\alpha = 5^\circ$, Reynolds number based on the airfoil chord $\Rey = 5 \times 10^4$ and Mach number $M = 0.4$. The investigation is carried out for two sweep angles $\Lambda = 20^\circ$ and $40^\circ$, and an unswept wing case ($\Lambda = 0^\circ$) is also considered for comparisons. Two different swept-wing configurations have been considered, the details of which are illustrated in figure \ref{fig:config_sketch}. The cases were designed as part of a larger project, and consider two different methods of installing swept wings in wind tunnels. They were not explicitly designed to verify or disprove the independence principle (which anyway is not strictly valid for non-laminar flows), although the simulations do provide useful insights in this aspect, as discussed in the next section. The free-stream speed $Q_\infty = 1$ is kept constant regardless of the sweep angle for both the configurations. Hence, the velocity components in the directions perpendicular and parallel to the leading edge of the wing vary with the sweep angle according to $U_\infty = Q_\infty \cos\Lambda$ and $W_\infty = Q_\infty \sin\Lambda$, respectively. The main difference between the two swept-wing configurations considered is that, while configuration A is obtained by rotating the unswept wing about the $y'$ axis, configuration B is given by shearing the profile in the $z$ direction, corresponding to the typical aircraft design practice of specifying wing sections at different normal distances from the aircraft fuselage centreline. The leading-edge-perpendicular section (section A-A in figure \ref{fig:config_sketch}) is the standard NACA-0012 section for configuration A and a scaled (in the $x$ direction, so that the computed chord $c_\perp = c \times \cos \Lambda$) version of this for configuration B, while the section thickness ($d$) remains the same for the two configurations. Note that, here, $c$ is defined as the distance between the leading and the trailing edges of the reference NACA-0012 profile. Table \ref{t:cases} gives a summary of the numerical simulations considered in the investigation. The Reynolds number based on the airfoil thickness is $\Rey_{d} = 6.0\times 10^3$ for all the cases, while the Reynolds number based on $c_\perp$ varies with the sweep angle $\Lambda$ for the configuration B cases.
%

%
\begin{table}
  \begin{center}
\def~{\hphantom{0}}
\begin{tabular}{lccccc}
Case & Sweep ($\Lambda$)   & Computed chord ($c_\perp/c$)         & Airfoil thickness ($d/c$)  & $\Rey_{c_\perp}$      & $\Rey_{d}$       \\ [3pt]
S0   & $0^\circ$   & $1.0$          & $0.12$ & $5.0\times 10^4$ & $6.0\times 10^3$ \\
S20A & $20^\circ$  & $1.0$          & $0.12$ & $5.0\times 10^4$ & $6.0\times 10^3$ \\
S20B & $20^\circ$  & $\cos \Lambda$ & $0.12$ & $4.7\times 10^4$ & $6.0\times 10^3$ \\
S40A & $40^\circ$  & $1.0$          & $0.12$ & $5.0\times 10^4$ & $6.0\times 10^3$ \\
S40B & $40^\circ$  & $\cos \Lambda$ & $0.12$ & $3.8\times 10^4$ & $6.0\times 10^3$ 
\end{tabular}
\caption{Details of the computational study. For all the cases $\Rey = 5\times 10^4$, $M = 0.4$ and incidence $\alpha = 5^\circ$.}
\label{t:cases}    
\end{center}
\end{table} 

\begin{figure}
  \centerline{\includegraphics[trim = 0mm 0mm 0mm 0mm,clip,width=0.5\textwidth]{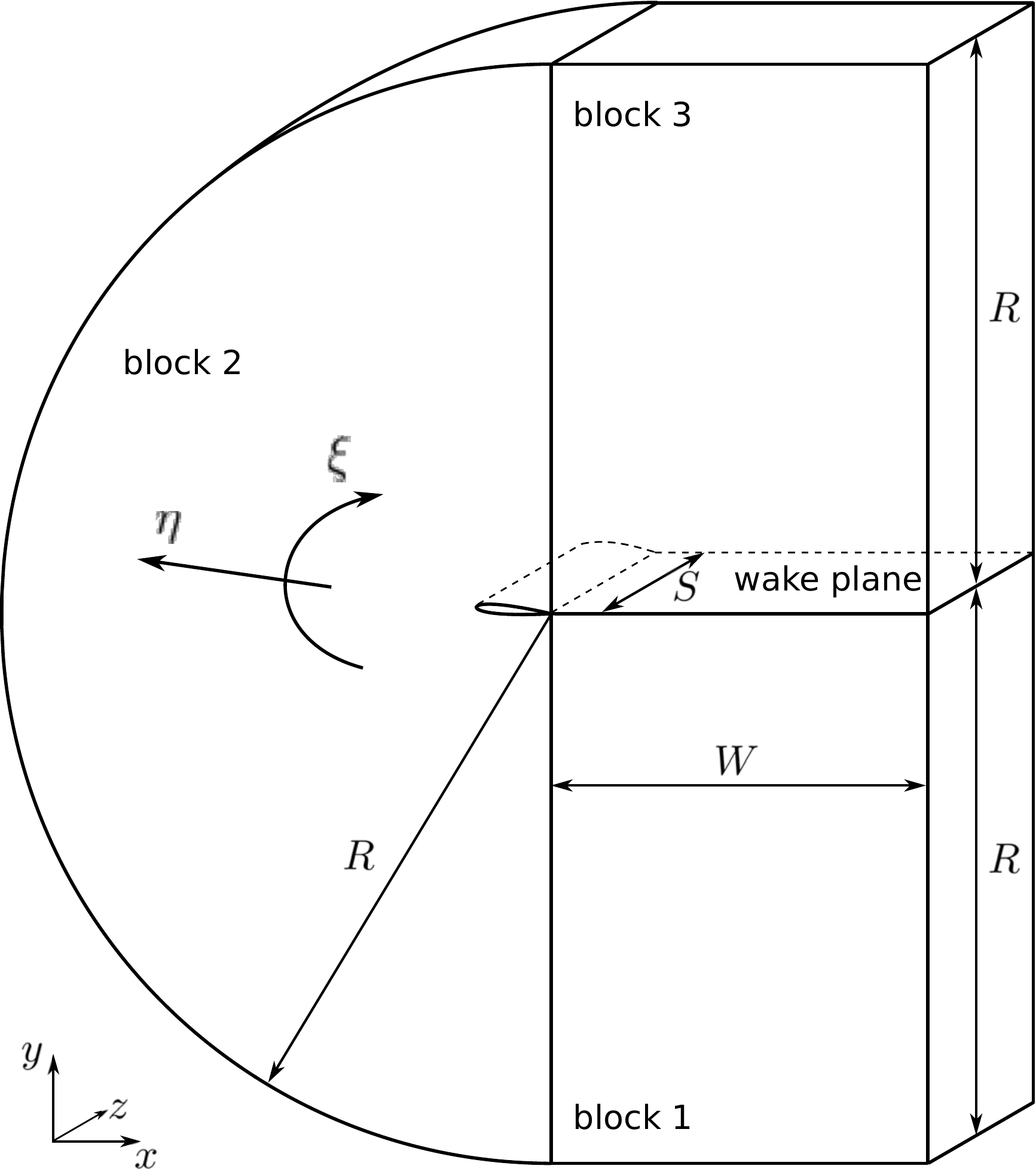}}
  \caption{Sketch of the computational domain used. Block 2 contains the NACA-0012 aerofoil, while blocks 1 and 3 meet at the wake plane.}
\label{fig:domain_sketch}
\end{figure}

A sketch of the computational domain used for the numerical simulations is shown in figure \ref{fig:domain_sketch}. The domain is divided into three blocks, with interface boundary conditions between neighbouring blocks, and its dimensions are $R = 7.3$, $W = 5$ and $S = 0.4$ for all the simulations. The NACA-0012 airfoil profile has a sharp trailing edge imposed (by extending and rescaling the profile slightly, as in \citet{jones08}). Blocks 1 and 3 both contain the wake plane, which should be identical up to machine accuracy. To prevent any possible divergence of these solutions, the averaging that is applied at the trailing edge is extended along the wake plane, where it serves only to sychronise the two solutions. The numerical simulations were carried out using characteristic conditions at all the computational domain boundaries, in order to minimize wave reflections. In particular, a zonal characteristic boundary condition \cite{zonalbc} is applied over a distance $L_{zonal} \approx 0.85c$ near the outflow boundary of blocks 1 and 3, using $61$ grid points. A standard characteristic condition \cite{thomson1,thomson2} is applied at the rest of the boundaries, where, in addition, outgoing characteristics are integrated over time and superimposed on the fixed free stream condition. The airfoil is modelled using a no-slip, isothermal boundary condition, with the wall temperature equal to the free stream temperature.

\begin{table}
  \begin{center}
\def~{\hphantom{0}}
\begin{tabular}{lccccccccccc}
Case    & R/c   & W/c   & S/c   & N$_{foil}$ & N$_{wake}$ & N$_{\xi}$ & N$_{\eta}$ & N$_z$ & $\Delta y^+_{min}$ & $\Delta x^+_{max}$ & $\Delta z^+_{max}$ \\ [3pt]
S0 & $7.3$ & $5.0$ & $0.4$ & $1799$     & $1602$     & $3401$     & $692$      & $240$ & $0.92$              & $3.46$        & $4.92$                        \\
\citet{jones_sandham08} & $7.3$ & $5.0$ & $0.2$ & $1066$     & $1506$     & $2570$     & $692$      & $96$ & $1.0$              & $3.36$        & $6.49$  \\
S20A & $7.3$ & $5.0$ & $0.4$ & $1799$     & $1602$     & $3401$     & $692$      & $240$ & $0.92$              & $3.45$        & $4.92$                      \\
S20B & $7.3$ & $5.0$ & $0.4$ & $1799$     & $1602$     & $3401$     & $692$      & $240$ & $0.84$              & $2.80$        & $4.48$                      \\
S40A & $7.3$ & $5.0$ & $0.4$ & $1799$     & $1602$     & $3401$     & $692$      & $240$ & $0.92$              & $3.43$        & $4.97$                      \\
S40B & $7.3$ & $5.0$ & $0.4$ & $1799$     & $1602$     & $3401$     & $692$      & $240$ & -              & -        & -                      
                      \\
\end{tabular}
\caption{Details of the computational study.}
\label{t:grid}    
\end{center}
\end{table} 
A summary of the features of the computational grid employed in this study is reported in table \ref{t:grid}. The same grid is used for all the numerical simulations and was chosen based on a grid convergence study carried out by \citet{jones_sandham08} for the same un-swept wing configuration analysed in this work. Compared to the computational grid used by \citet{jones_sandham08}, the grid used here has been further refined in the $x$ and $z$ directions, especially around the airfoil section. Based on the values of $\Delta y^+_{min}$, $\Delta x^+_{max}$ and $\Delta z^+_{max}$, which were calculated at the position of the skin friction peak after the laminar-turbulent transition on the suction side of the wing, the same grid is deemed appropriate also for the swept-wing cases. The grid resolution in wall units is not reported for the S40B case, since, as will be shown later, the flow over the wing is not turbulent.

%
\begin{figure}
  \centerline{\includegraphics[trim = 0mm 0mm 0mm 0mm,clip,width=1.0\textwidth]{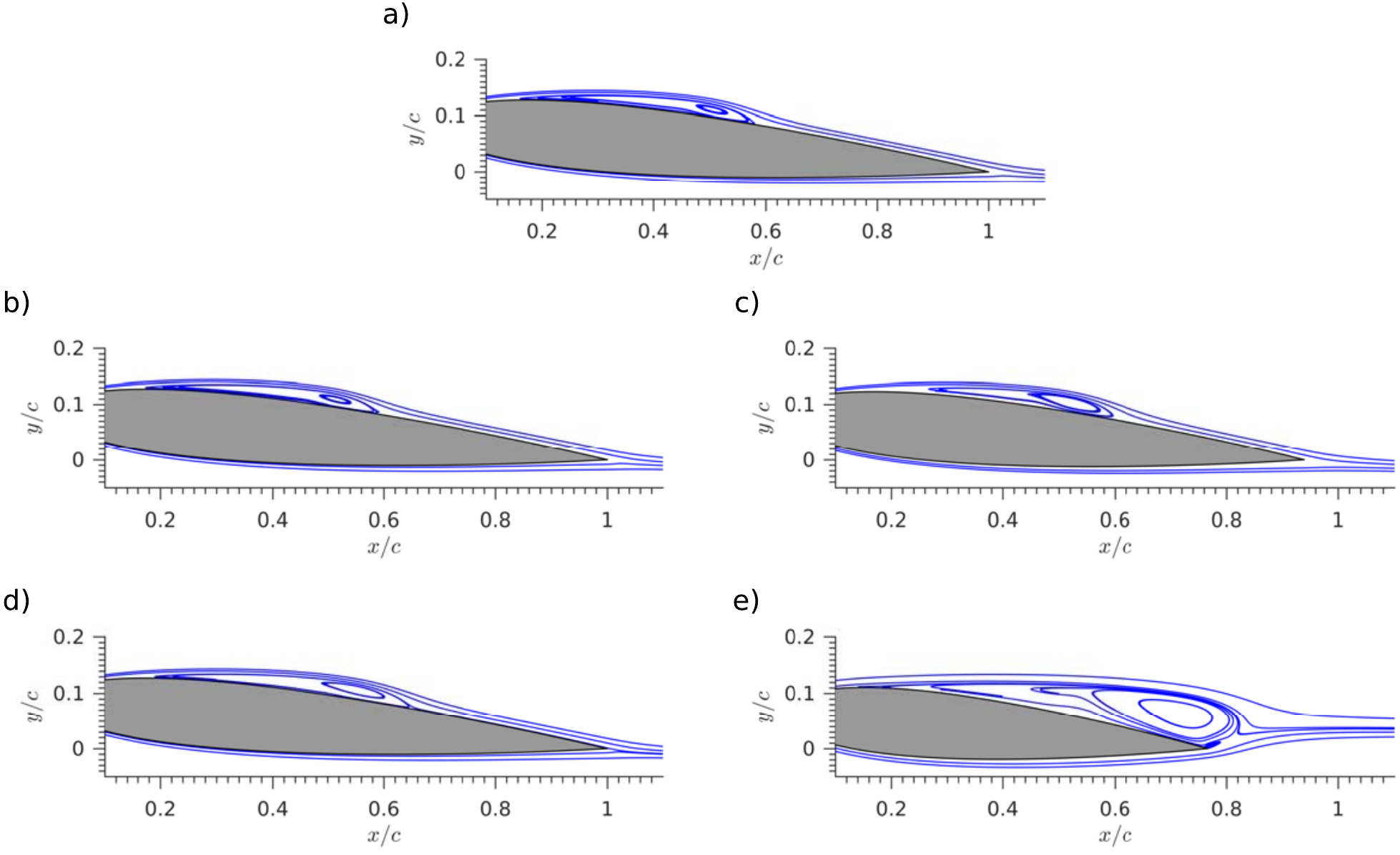}}
  \caption{Streamlines for the time- and span-averaged flow over the airfoil. a) case S0, b) case S20A, c) case S20B, d) case S40A, e) case S40B.}
\label{fig:streamlines}
\end{figure} 
\begin{figure}
  \centerline{\includegraphics[trim = 0mm 0mm 0mm 0mm,clip,width=1.0\textwidth]{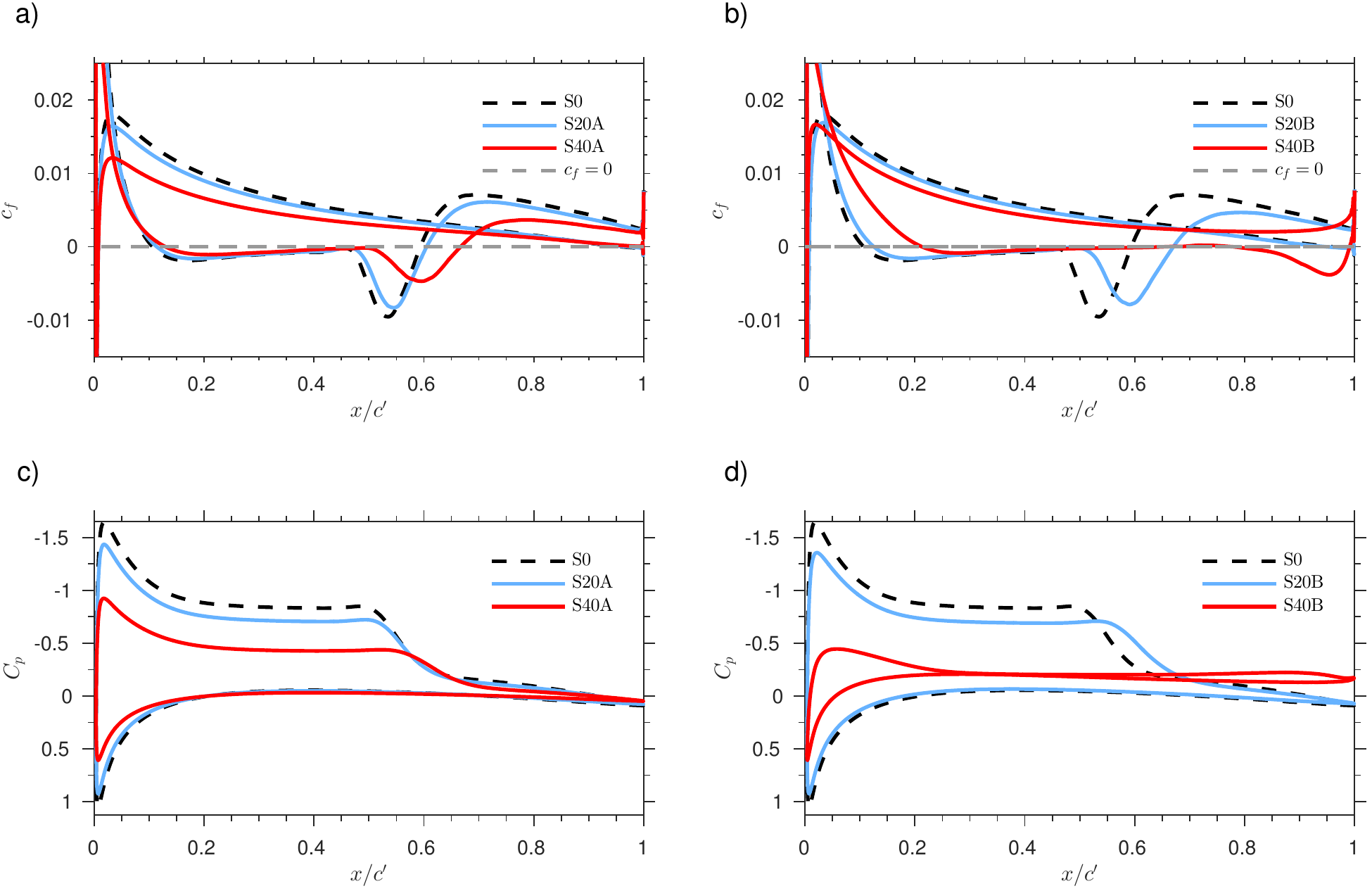}}
  \caption{a)-b) Skin friction ($c_f$) and c)-d) pressure coefficient ($C_p$) distributions over the airfoil's surface.}
\label{fig:coeficients}
\end{figure} 
\section{Mean flow features} \label{sec:mean_flow}
The mean flows analysed in this section are calculated by taking time-averages over an interval of $19$ dimensionless time units (equivalent to at least $15$ wing flow-pasts, depending on the case considered) and averaging over the $z$ direction. The averaging procedure was started after the 3D simulations were allowed to settle for $20$ time units. 

A view of the mean flow features is provided in figure \ref{fig:streamlines}, showing the distribution of mean streamlines around the airfoil for all the cases considered. Figure \ref{fig:streamlines}(a) shows that in the un-swept case a separation bubble forms on the suction side of the airfoil, extending for about half the airfoil chord, while the flow is attached everywhere else. Similar flow characteristics can be observed in figures \ref{fig:streamlines}(b)-(d) for cases S20A, S40A and S20B , with slight differences in the structure of the separation bubbles. On the other hand, the mean flow around the airfoil appears radically different for case S40B in figure \ref{fig:streamlines}(d). In this case, the boundary layer on the suction side does not reattach after separation, so that a large region of separated flow dominates the flow over the top of the airfoil. 

The skin friction distributions shown in figure \ref{fig:coeficients}(a) indicate that the laminar boundary layer on the airfoil's suction side separates near $x/c = 0.12$ for cases S0 and S20A and near $x/c = 0.15$ for case S40A. Following laminar boundary layer separation, a laminar-turbulent transition process takes place inside the bubble, the late stages of which lead to the sharp skin friction rise notable near the half chord position. Figure \ref{fig:coeficients}(b) shows the skin friction distributions for cases S20B and S40B, with case S0 as reference. Turbulent reattachment of the flow is delayed considerably for case S20B (with respect to the unswept case), leading to a separation bubble length of $0.5$ times the airfoil chord, while the flow never reattaches to the surface of the airfoil for case S40B. The effect of sweep on the pressure coefficient ($C_p$) distribution on the airfoil's surface is shown in figures \ref{fig:coeficients}(c) and \ref{fig:coeficients}(d) for configurations A and B, respectively. The magnitude of the pressure coefficient decreases as the sweep angle is increased, leading, as will be shown later, to the expected decrease of lift. The $C_p$ distribution for case S40B in figure \ref{fig:coeficients}(d) indicates stall.

\begin{table}
  \begin{center}
\def~{\hphantom{0}}
\begin{tabular}{lcccccccc}
Case & NACA     & $\Rey$           & $M$    & $C_L$  & $C_D$ & $x_s$ & $x_t$ & $x_r$  \\ [3pt]
S0   & $0012$   & $5.0\times 10^4$ & $0.40$ & 0.680   & 0.037   & 0.06   & 0.44  & 0.53 \\
S20A & $0012$   & $4.7\times 10^4$ & $0.38$ & 0.677   & 0.038   & 0.06   & 0.47  & 0.58 \\
S40A & $0012$   & $3.8\times 10^4$ & $0.31$ & 0.608  & 0.051   & 0.08   & 0.62  &  0.79 \\
S20B & $0013$   & $4.4\times 10^4$ & $0.38$ & 0.627  & 0.050   & 0.08   & 0.56  & 0.71 \\
S40B & $0016$   & $2.9\times 10^4$ & $0.31$ & -0.022  & 0.057   & 0.19   & 1.00 & 1.0 
\end{tabular}
\caption{Results from XFoil based on the flow and geometry normal to leading edge, using a critical N factor of 13. The airfoil section is taken as the closest NACA 4-digit section.}
\label{t:xfoil}    
\end{center}
\end{table} 

The effects of sweep and Reynolds number can be more completely understood by combining the results shown in \ref{fig:coeficients} with predictions from the XFoil package (https://web.mit.edu/drela/Public/web/xfoil/), which is capable of computing transitional separation bubbles of the kind seen in the DNS (reference e.g. Drela-Giles maybe). XFoil doesn't include sweep effects, but does allow the effect of Reynolds number to be assessed. For these XFoil calculations the critical amplification factor was set to $N=13$ to optimise bubble length predictions compared to the DNS. Table 2 lists the equivalent XFoil results, considering only the flow perpendicular to the leading edge. The table includes the equivalent unswept airfoil, based on closest NACA airfoil section, the lift and drag coefficients well as the separation ($x_s$), transition ($x_t$) and reattachment ($x_r$) points. For both cases the separation bubble location and trend is in good agreement with \ref{fig:coeficients} including the rearward movement of the reattachment and transition points (transition can be roughly matched to the minimum skin friction point in the DNS). Thus the small variations in the bubble are likely to be Reynolds number effects rather than sweep. The change in level of the pressure plateau is partially a Reynolds number effect, but mainly an effect of sweep. For Case B, XFoil shows massive separation, which is also in agreement with the DNS.  

%

The ability of XFoil to predict the behaviour of certain features of the separation bubbles for equivalent 2D cases is empirical evidence that an independence principle may be approximately valid, despite certain conditions not being fulfilled. For swept flow it is possible to show that the introduction of a spanwise velocity component does not affect the flow in the direction normal to the wing's leading edge, as the equations of motion can be decomposed into two independent terms, one tangential and one perpendicular to the wing leading edge. This independence principle is exact only for laminar, incompressible flows \citep{hetsch09a} and for laminar compressible flows without viscosity. It cannot even be assumed to hold in the laminar region near the leading edge in incompressible flow, since turbulent flow region downstream can change the potential flow via a boundary layer displacement effect. As previously mentioned, the present cases were not designed explicitly to assess the independence principle, but it is useful to document the lift and drag results in terms of this framework. 
 
\begin{figure}
  \centerline{\includegraphics[trim = 0mm 0mm 0mm 0mm,clip,width=0.9\textwidth]{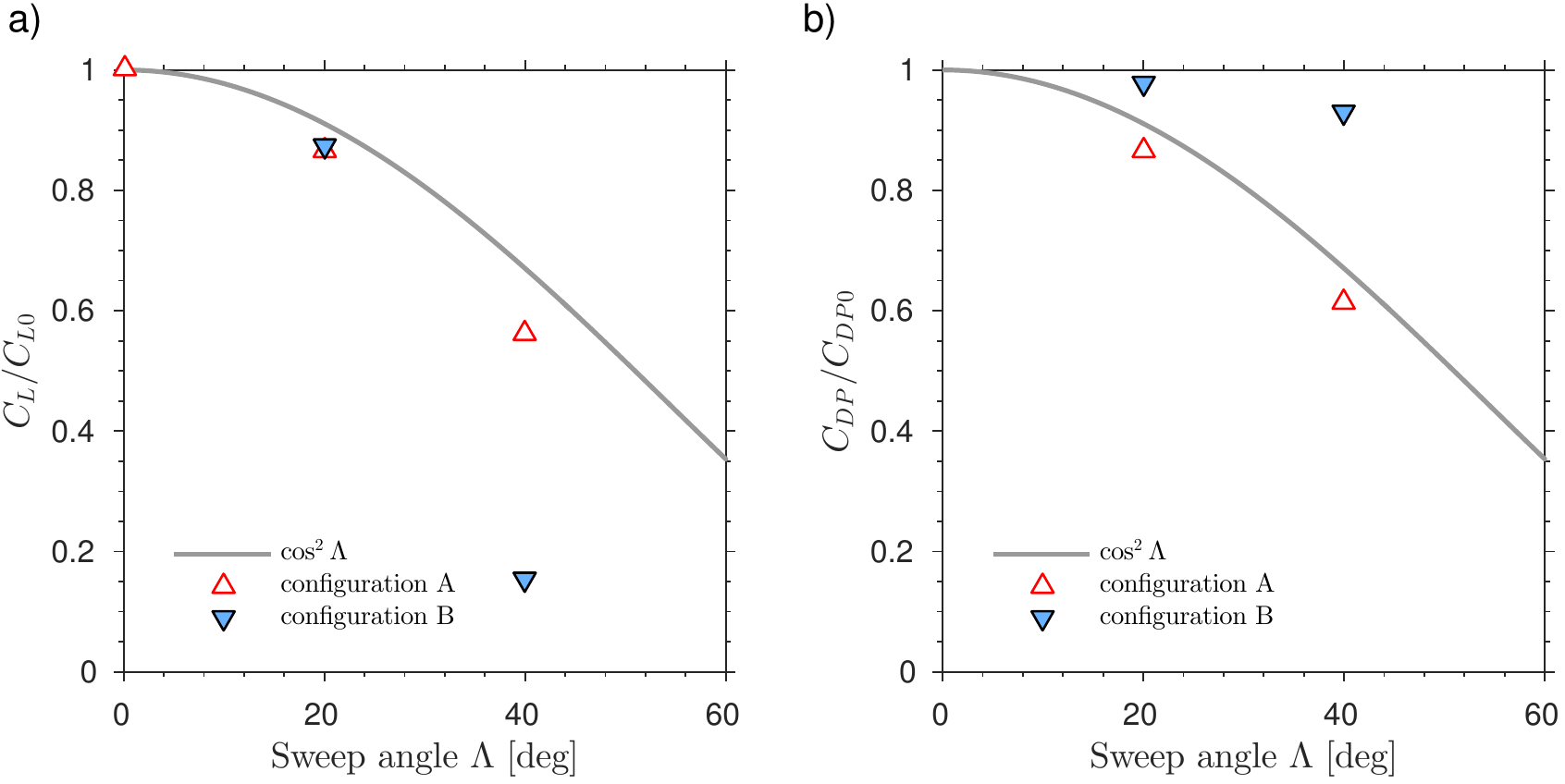}}
  \caption{Evaluation of the independence principle for the calculation of a) lift coefficient ($C_L$) and b) pressure drag coefficient ($C_{DP}$).}
\label{fig:simple_sweep}
\end{figure}

In the set up considered in this work, the reduction of normal velocity component $U_\infty$ with sweep angle $\Lambda$ (recall that $U_\infty = Q_\infty \cos \Lambda$, with $Q_\infty$ constant) leads to a reduction of lift and pressure drag, due to the reduced leading-edge-normal dynamic pressure. Considering only the part due to pressure differences and neglecting Reynolds number effects, the lift and pressure drag coefficients are given by $C_L = C_{L0} \cos^2\Lambda$ and $C_{DP} = C_{DP0} \cos^2\Lambda$, where $C_{L0} = 6.17\times10^{-1}$ and $C_{DP0} = 2.73\times10^{-2}$ are, respectively, the lift and drag coefficients of the unswept wing case S0. Figures \ref{fig:simple_sweep}(a) and \ref{fig:simple_sweep}(b) show, respectively, $C_L/C_{L0}$ and $C_{DP}/C_{DP0}$ as a function of sweep angle. The results for configuration A agree closely with the theoretical predictions, indicating that the lift-to-drag ratio is kept nearly constant as the sweep angle is increased. In Configuration B the effect of thickness would also need to be taken into consideration.
\begin{figure}
  \centerline{\includegraphics[trim = 0mm 0mm 0mm 0mm,clip,width=0.9\textwidth]{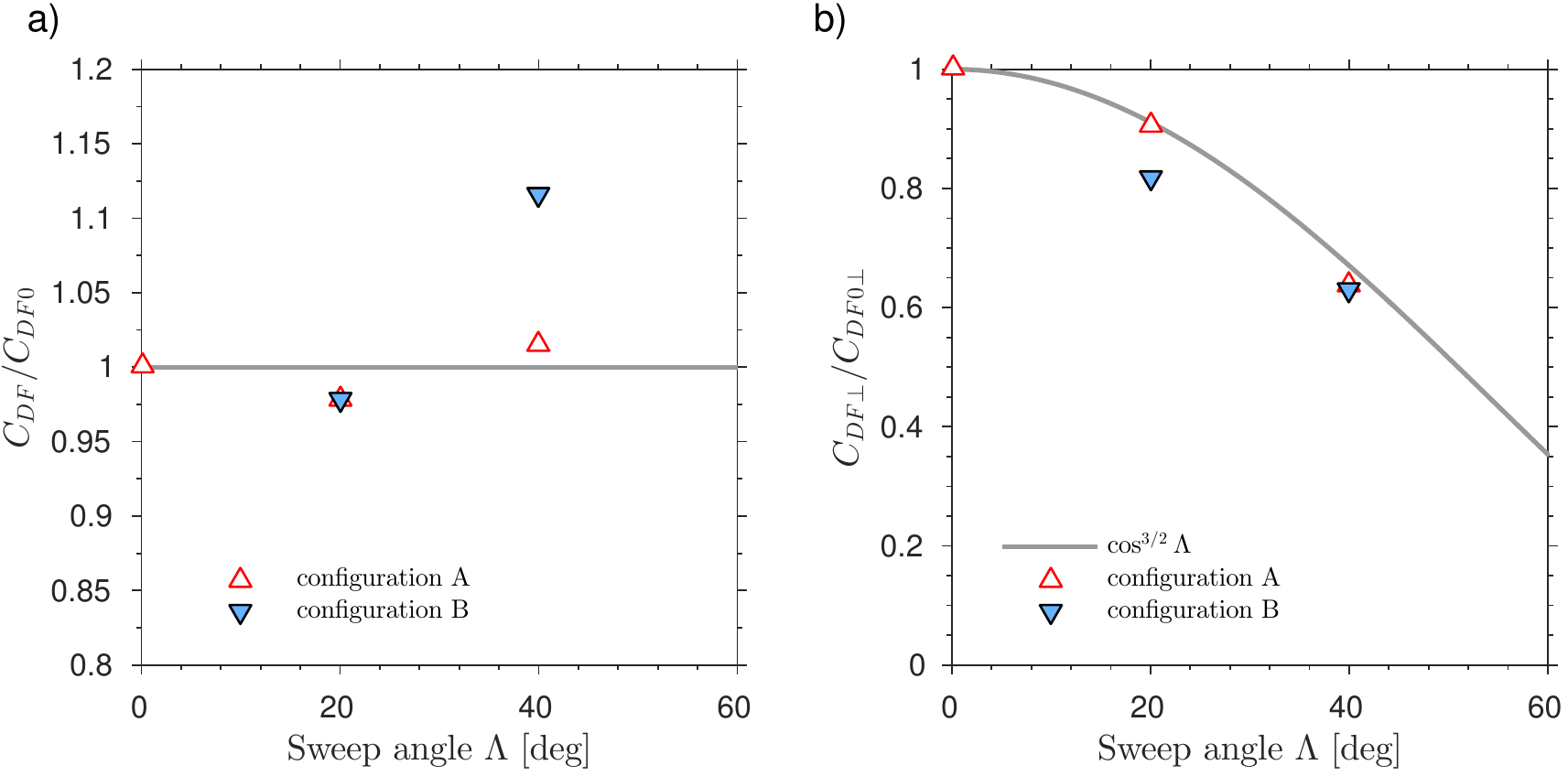}}
  \caption{Evaluation of the independence principle for the calculation of a) total friction drag coefficient ($C_{DF}$) and b) friction drag coefficient perpendicular to the leading edge ($C_{DF\perp}$).}
\label{fig:simple_sweep_cf}
\end{figure} 

A separate approach is needed for the friction drag, since this follows a different behaviour. One possible way to test the independence principle for the viscous forces is to look at the behaviour of the total friction drag coefficient as sweep increases, shown in figure \ref{fig:simple_sweep_cf}a). Since, for configuration A, sweep is introduced by simply rotating the airfoil around the vertical axis, if the independence principle holds, the total friction drag (per unit span) should remain constant. This can be seen to be nearly the case for configuration A, while the same behaviour is not observed for the cases belonging to configuration B. Another way to test the independence principle is to monitor how the friction drag in the direction perpendicular to the leading edge varies with sweep angle. Since in the cases analysed here, the introduction of sweep is accompanied by a decrease of the free stream velocity component in the direction perpendicular to the leading edge (and hence Reynolds number $\Rey_{U_\infty,c}$), a variation of the friction drag in this direction is also expected. In laminar flat plate boundary layers, the friction drag and the Reynolds number are related by the well known formula $C_{DF} = \kappa \Rey_{U_\infty,c}^{-1/2}$. Assuming that this is a good approximation for the flows investigated in this work, and considering that here $U_\infty = Q_\infty \cos \Lambda$, it is easy to show that from the independence principle it follows that the friction drag coefficient perpendicular to the leading edge should vary with sweep angle as $C_{DF\perp} = C_{DF0\perp} \cos^{3/2} \Lambda$, where $C_{DF0} = 4.6\times10^{-3}$ is the friction drag coefficient of the unswept configuration. Figure \ref{fig:simple_sweep_cf}b) shows that this is, in fact, the case for the cases belonging to configuration A. These results indicate that the friction drag variation with Reynolds number is well approximated by laminar flat plate boundary layer theory and thus the effects of sweep on skin friction drag can be accounted for. Note, however, that here we are dealing with flows with a large laminar proportion, unlike in the work of \citet{wygnanski14} where the independence principle was investigated in the turbulent flow regime.

\section{Unsteady flow features}
\label{sec:unsteady_features}
\begin{figure}
\centerline{\includegraphics[trim = 0mm 0mm 0mm 0mm,clip,width=1.0\textwidth]{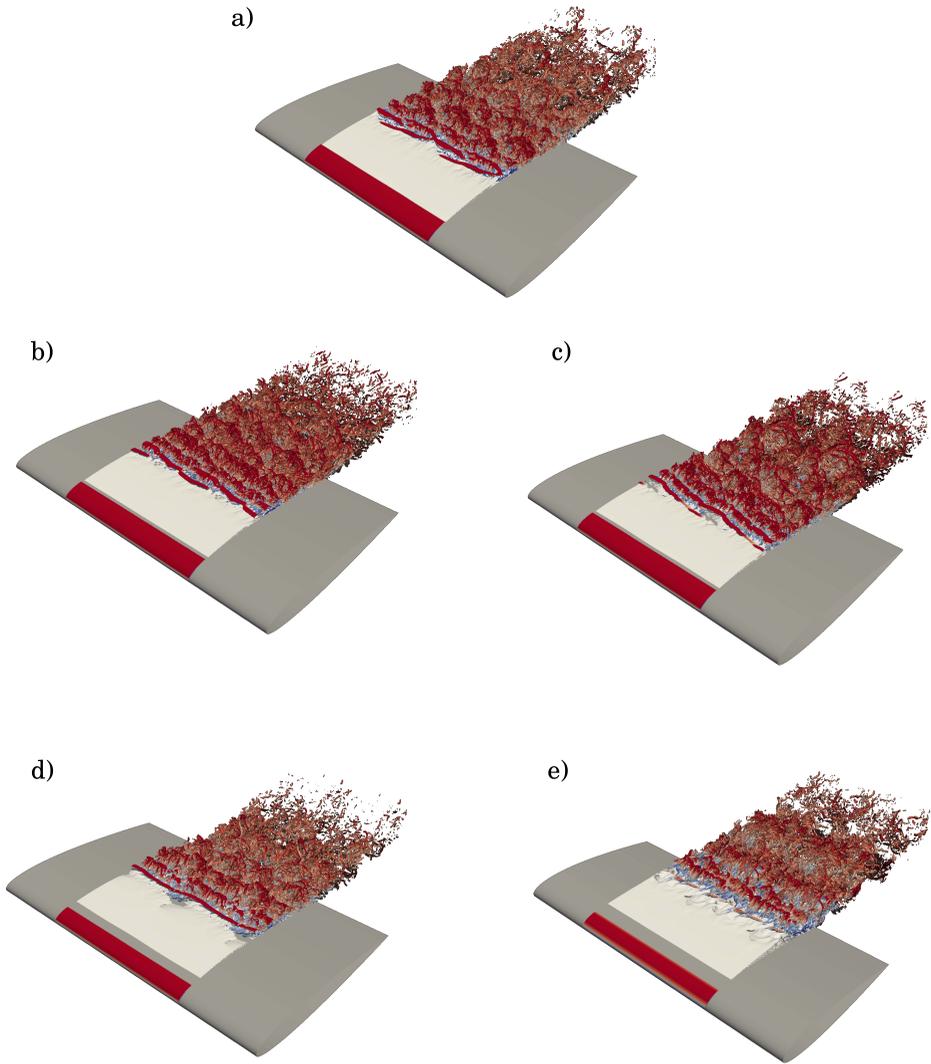}}
  \caption{Isosurfaces of the second invariant of the velocity gradient tensor ($Q = 200$) showing the dominant vortical structures in the transitional and turbulent regions of the flow. a) case S0, b) case S20A, c) case S20B, d) case S40A and e) case S40B. Two spans of the computational domain are shown.}
\label{fig:isosurfaces}
\end{figure} 
Figure \ref{fig:isosurfaces} shows the flow structures in the transitional flow over the suction side of the airfoil through isosurfaces of the second invariant of the velocity gradient tensor (Q-criterion) for all the cases considered, including two spans of the computational domain (0.8$c$ in total) to help interpret the flow features. The dominant flow structures in the transitional separation bubble for case S0, shown in figure \ref{fig:isosurfaces}(a), indicate that instabilities of the detached shear layer drive the initial stages of the laminar-turbulent transition process. Similar features can be observed for the cases S20A and S40A in figures \ref{fig:isosurfaces}(b), and \ref{fig:isosurfaces}(d), respectively, but with some significant changes in the coherence and organisation of the transitional structures, as discussed below. For case S20B (\ref{fig:isosurfaces}(c)), and in particular for case S40B (\ref{fig:isosurfaces}(e)), the transition process is delayed, consistent with the observation that the independence principle does not apply to these cases. 

Several changes in flow structure are observed with the introduction of sweep. Firstly, comparing figures \ref{fig:isosurfaces}(b), and \ref{fig:isosurfaces}(c) with figure \ref{fig:isosurfaces}(a) we see that the introduction of modest sweep leads to structures that are more coherent and more two-dimensional, with the orientation being perpendicular to the freestream direction. The increased coherence is maintained at the higher sweep angle (figures \ref{fig:isosurfaces}(d), and \ref{fig:isosurfaces}(e)), but in these cases the orientation of the structures is parallel to the leading edge, with a dominant mode of $k_z = 0$. The zero spanwise wavenumber for cases S40A and S40B means that the wave vector forms a $40^\circ$ angle with the free-stream flow direction. In the remainder of the paper, modes with zero spanwise wavenumber will be referred to as two-dimensional modes, while modes with non-zero spanwise wavenumber will be referred to as three-dimensional modes. It should be noted that no external disturbances have been imposed in the numerical simulations carried out in this work, so that the laminar-turbulent transition is self-sustained and cannot be attributed solely to the development of convective instabilities. 

The appearance of increased coherence was unexpected and the reasons will be discussed in the following paragraphs. The largest spanwise wavenumber contained in the computational box is $k_z=2\pi/0.4=15.71$ which, although improved by a factor of two relative to previous simulations \citep{jones_sandham08} of the same airfoil, sets a limit on the largest structures. The result is that the angles of any coherent oblique structures are limited to a set of discrete values. As an example, for the approximate streamwise structure spacing of $\lambda=0.05$ observed in the S20A and S20B cases, the possible angles $\theta$ are $7.13^\circ$, $14.0^\circ$, $20.6^\circ$ for increasing spanwise wavenumber (according to $\tan \theta=k_z\lambda/(2\pi)$). The angular resolution, and hence the spanwise domain size, is thus sufficient for the conclusions that structures are aligned with the flow direction for the moderately swept case, as well as the emergence of 2D structures in the highly swept cases. 
\begin{figure}
  \centerline{\includegraphics[trim = 0mm 0mm 0mm 0mm,clip,width=0.8\textwidth]{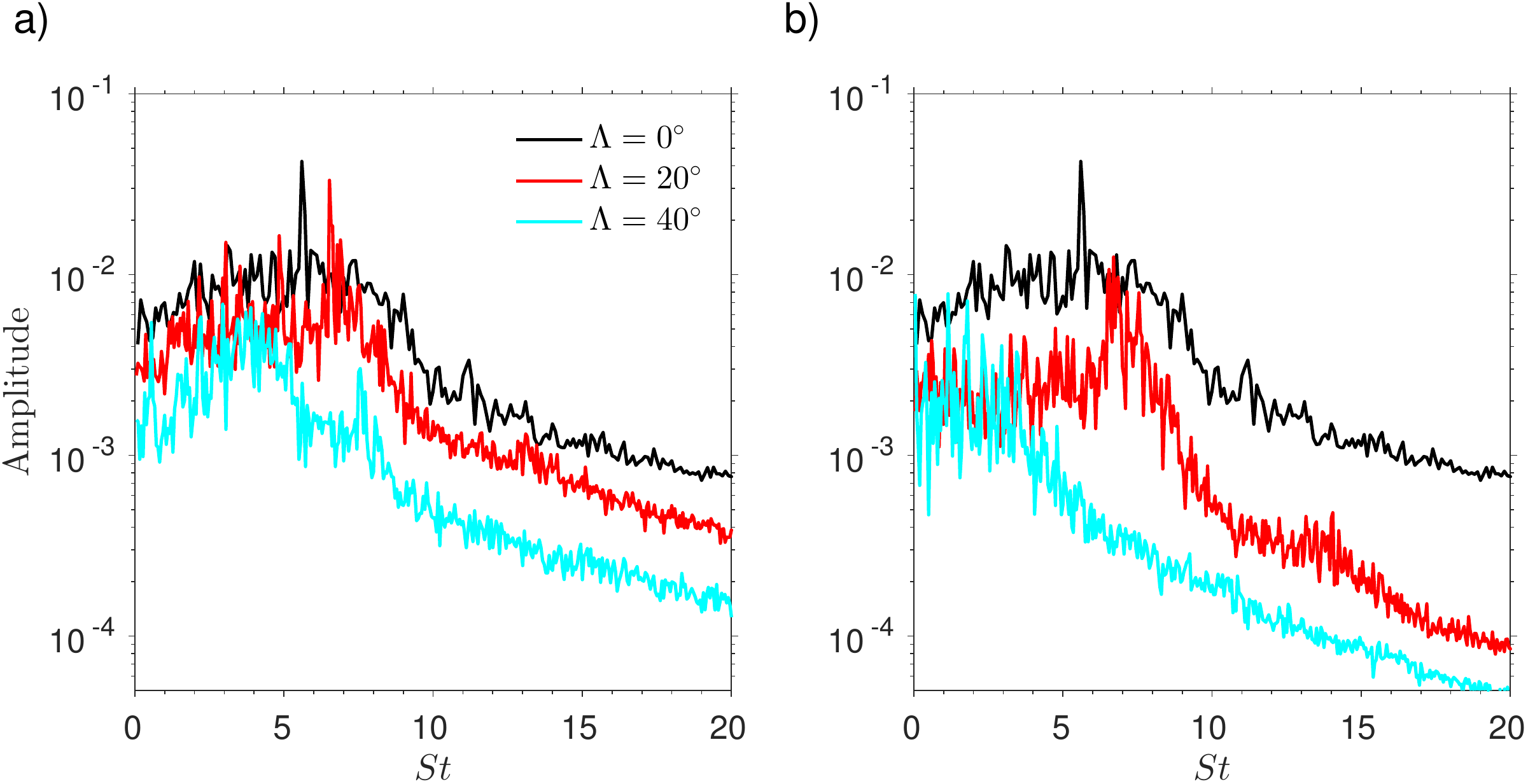}}
  \caption{Span-averaged temporal Fourier spectra for the wall pressure at the location of the initial vortex shedding location. a) configuration A, b) configuration B.}
\label{fig:Fourier_spectra}
\end{figure} 
\begin{figure}
\centerline{\includegraphics[trim = 0mm 0mm 0mm 0mm,clip,width=1.0\textwidth]{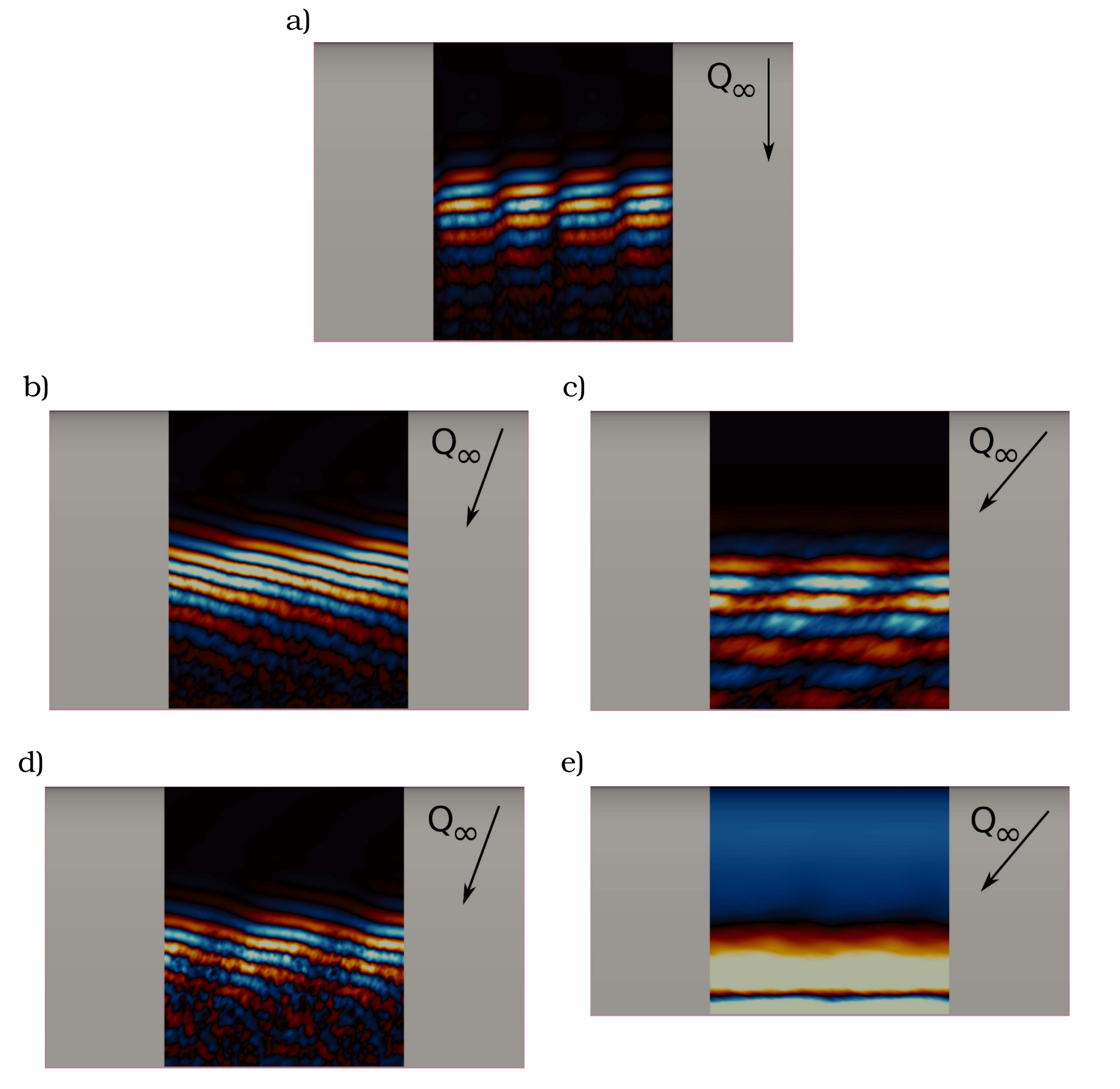}}
  \caption{Contours of the real part of the wall pressure Fourier coefficients on the suction side at the dominant frequencies. a) case S0 ($St=5.6$), b) case S20A($St=6.53$) , c) case S40A($St=7.56$), d) case S20B($St=6.7$) and e) case S40B($St=1.15$). Two spans of the computational domain are shown.}
\label{fig:Fourier_modes}
\end{figure} 

The dominant flow structures can also be identified from their wall pressure imprint. Pressure time series were accumulated for all spanwise positions over the whole surface of the wing. Span-averaged spectra
are shown on figure \ref{fig:Fourier_spectra}, at a wall position under the initial vortex shedding location in each case. From these plots we can identify dominant frequencies for further study. In particular we can extract the spanwise structure by Fourier transforming the entire surface time history. Figure \ref{fig:Fourier_modes} shows contours of the real part of the Fourier coefficients over the suction side for the largest amplitude modes. This technique allows us to observe the three-dimensional structure of the most energetic modes. In the unswept case the separation bubble mode with Strouhal number $St \approx 5.6$ is shown. It has a structure typical of oblique-mode breakdown \citep{oblique_breakdown}. For $\Lambda = 20^\circ$ the Strouhal number of the highest amplitude mode is $St \approx 6.53$ for case S20A and $St \approx 6.7$ for case S20B, with oblique waves of one orientation (perpendicular to the free stream) seen to be dominant. For case S40A vortex shedding is excited by a broad range of 2D instability modes growing in the bubble, the most important of which has a frequency of $St \approx 3.0$. In this case a peak also appears at a Strouhal number $St \approx 7.56$ (see the blue line in figure \ref{fig:Fourier_spectra}) that is associated with an oblique mode with $k_z = +15.71$ (note that $z$ refers to a reference frame attached to the wing, with $x$ being the chordwise direction). The initial stages of laminar-turbulent transition are also driven by a range of 2D modes for case S40B, with peaks at $St = 1.15$ and $St = 1.8$. 

The Fourier decomposition highlights the underlying changes in flow structure that were already visible in the instantaneous snapshots, moving from oblique breakdown in the unswept case, to one-sided oblique modes oriented perpendicular to the free steam at moderate sweep angles, and spanwise-coherent structures at high sweep angles. Wider-domain calculations would be needed to explore the limits of the coherent structures. The objective of the rest of this paper is to explore reasons for the changes in flow structure, for which we will only consider the configuration A cases.
%
\section{Dynamic mode decomposition and global stability analysis}\label{sec:dmd_global}
\subsection{Separation bubble dynamics}
\begin{figure}
 \centerline{\includegraphics[trim = 0mm 0mm 0mm 0mm,clip,width=1.0\textwidth]{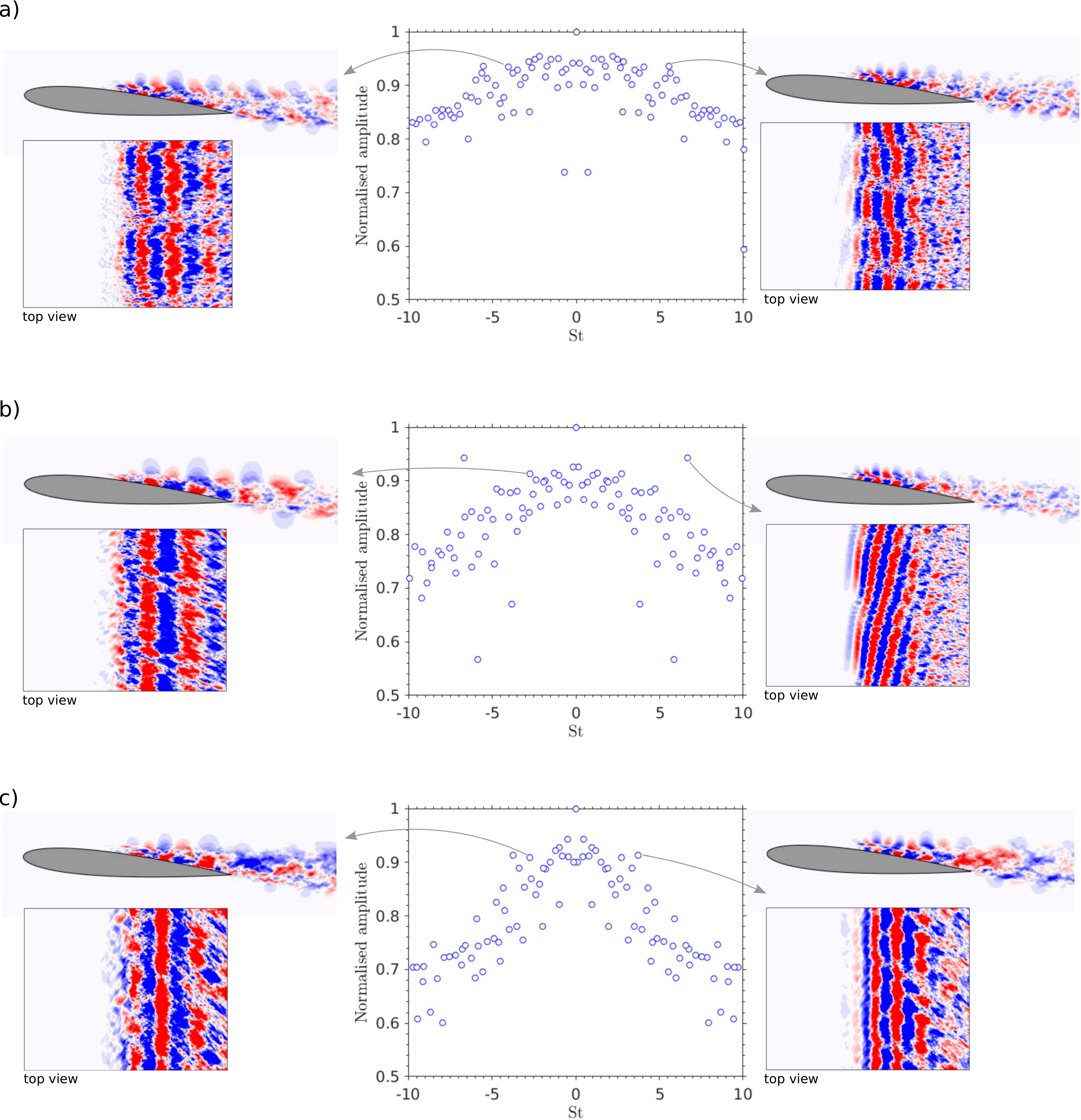}}
  \caption{Dynamic mode decomposition spectra and contours of the real part of the $u$-velocity for some of the most energetic DMD modes. a) case S0), b) case S20A and c) case S40A. Two spans of the computational domain are shown.}
\label{fig:DMD_All}
\end{figure} 
The dominant structures observed in the transitional separation bubble for cases S0, S20A and S40A were also investigated using Dynamic Mode Decomposition (DMD) \citep{schmid10}, with the objective of adding to the insights obtained from the instantaneous snapshots and wall pressure spectral analysis of the previous section. Here we use the streaming DMD algorithm proposed by \citet{hemati14}, which drastically reduces the memory requirements for the computation of DMD modes. In all the three cases analysed, the DMD analysis is carried out using an $x-y$ slice placed at $z=0.2$, combined with an $x-z$ plane on the upper side of the airfoil placed at a distance of about $6\times10^{-3}$ chords from the surface. The analysis is carried out over a time period of $\tau = 50$ chord-flow-through times, using snapshots of the full conservative variables vector $q = [\rho, \rho u, \rho v, \rho w, \rho E]^T$ separated by a constant time step $\Delta t = 0.05$. Figures \ref{fig:DMD_All}(a), (b) and (c) show the DMD spectra, together with contours of the real part of the $u$-velocity mode functions for two of the most energetic modes for cases S0, S20A and S40A, respectively. The mode amplitudes were scaled with the amplitude of the mean flow mode (the most energetic mode), seen in the spectra of figure \ref{fig:DMD_All} at $St = 0$ and unit amplitude. While some of the DMD modes extracted correspond to dynamical features of the turbulent boundary layer, the analysis here is focused on the modes that relate to the separation bubble transition process and subsequent vortex shedding cycle. Despite the large sweep angles considered in this work, it is interesting to note that crossflow instabilities were not found amongst the dominant DMD modes (or in the previous Fourier analysis) and do not seem to play a significant role in these flows.

For case S0 in figure \ref{fig:DMD_All}(a) one of the dominant DMD modes occurs for a Strouhal number of $St = 5.56$. This frequency was also identified by the Fourier analysis in section \ref{sec:unsteady_features} and, as suggested by the associated eigenfunction on the right hand side of the figure, relates to a K-H instability of the detached shear layer above the recirculation bubble, with a spanwise structure consistent with that obtained from the Fourier analysis as a superposition of two equal and opposite oblique waves. An example of a 2D DMD mode for this case is shown on the left hand side of \ref{fig:DMD_All}(a); it is also of K-H type and has a frequency $St = 4.0$. Case S20A, shown in figure \ref{fig:DMD_All}(b), has a distinct peak in the DMD spectrum at $St = 6.66$, which again corresponds with the dominant structure identified in the Fourier analysis. The DMD eigenfunction (the right hand plot) indicates that this mode is an oblique K-H instability of the detached shear layer above the bubble, travelling in the positive spanwise ($z$) direction. Two-dimensional modes are also present in the laminar-turbulent transition process in this case, the most amplified of which is shown on the left hand side figure \ref{fig:DMD_All}(b) and has a frequency of $St = 2.74$. In contrast, the dominant K-H modes are all 2D for case S40. Figure \ref{fig:DMD_All}(c) shows that the S40A DMD spectrum has a peak at around $St = 3.0$ and the modes with $St = 2.7$ (left) and $St = 3.7$ (right) are both manifestations of a 2D K-H instability of the detached shear layer. The DMD analysis using more complete flow information confirms the general picture obtained from the wall pressure spectra, and additionally shows the presence of 2D modes for lower sweep angles. The reasons for this behaviour will be explored further when we consider the global instability characteristics of the various cases in the next subsection.

\begin{figure}
  \centerline{\includegraphics[trim = 0mm 0mm 0mm 0mm,clip,width=0.9\textwidth]{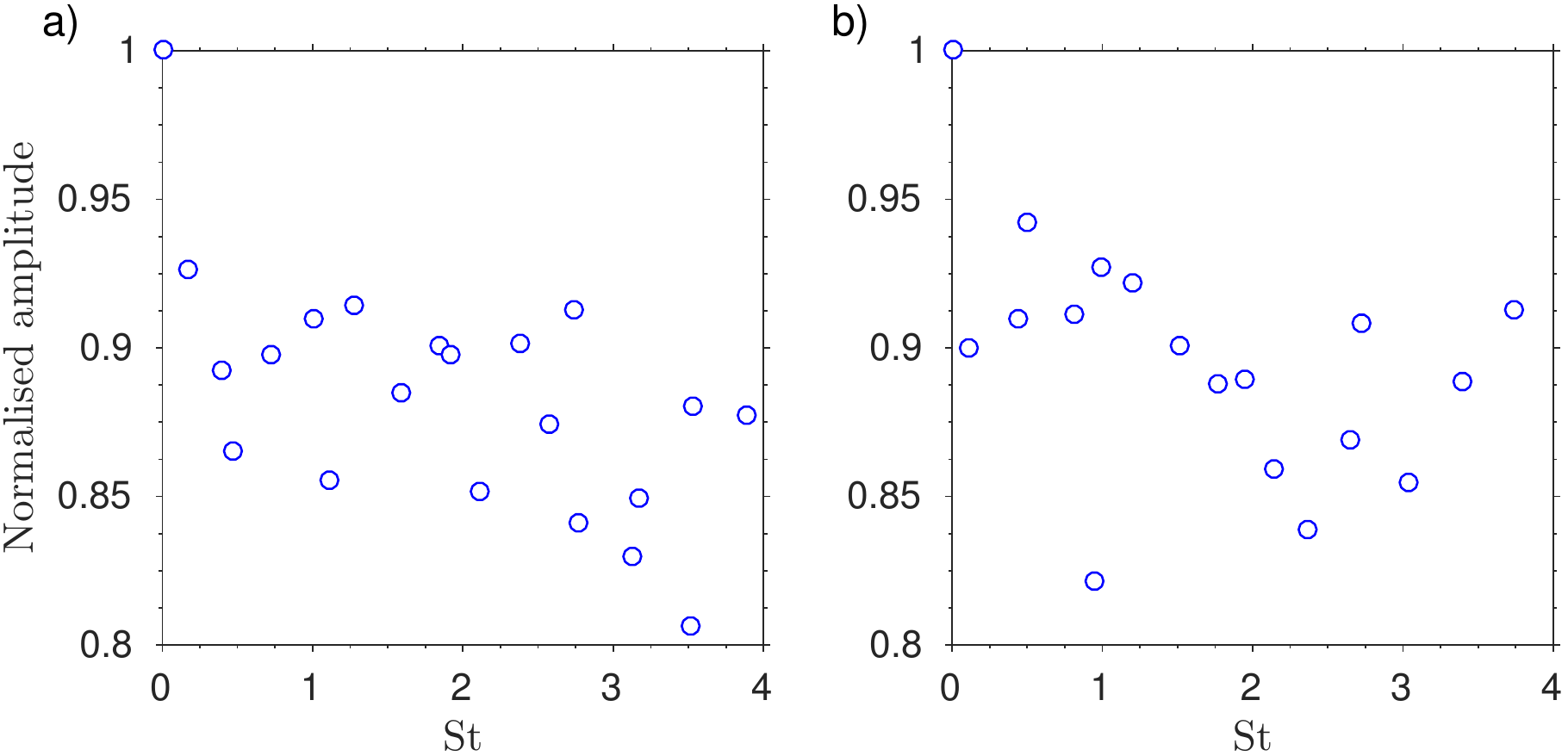}}
  \caption{Close-up view of the low frequency part of the DMD spectrum for: a) case S20A and b) case S40A.}
\label{fig:DMD_LowFreq}
\end{figure} 
\begin{figure}
  \centerline{\includegraphics[trim = 0mm 0mm 0mm 0mm,clip,,width=1.0\textwidth]{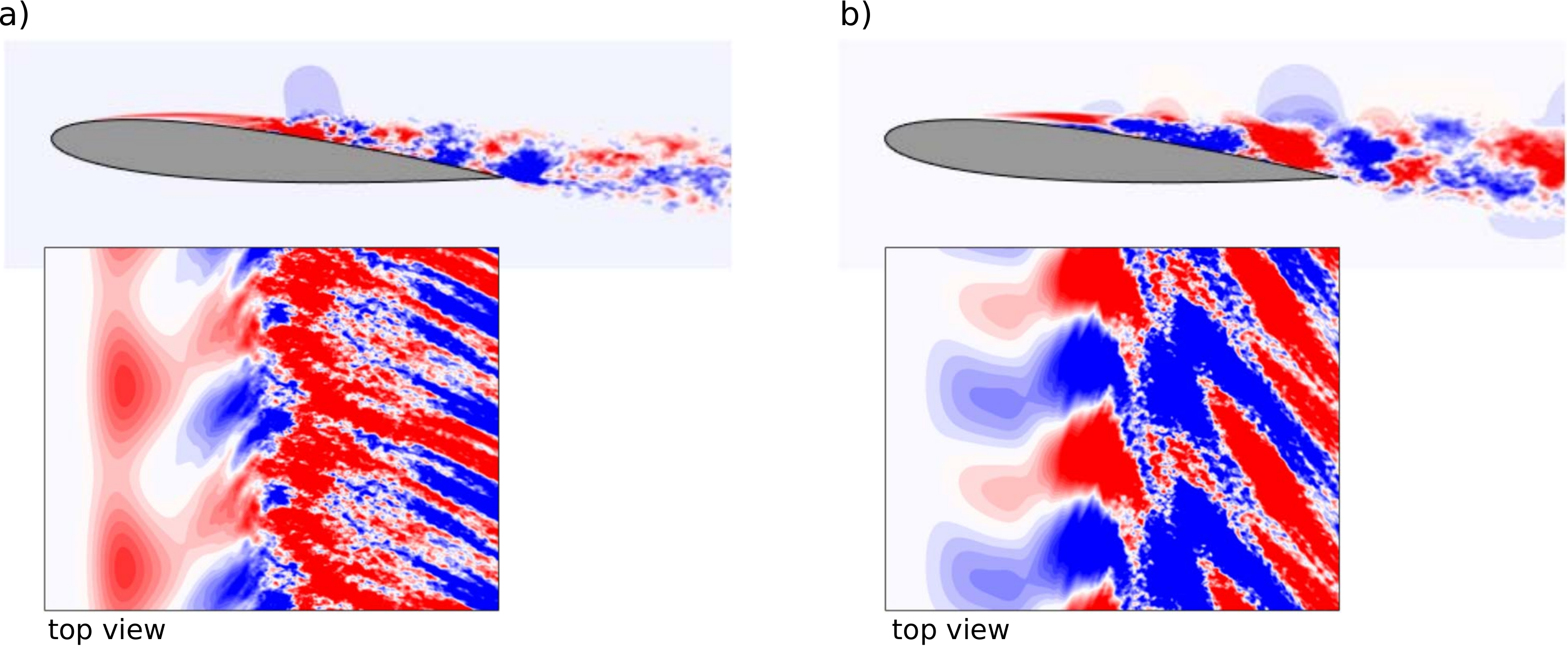}}
\caption{Low frequency bubble DMD modes, shown through contours of the real part of the $u$-velocity eigenfunction. a) case S20A, $St = 0.18$,  and b) case S40A, $St = 0.5$. Two spans of the computational domain are shown.}
\label{fig:low_frequency}
\end{figure}

In addition to the self-sustained K-H modes that lead to the characteristic vortex shedding shown in figure \ref{fig:isosurfaces}, a different kind of mode, characterised by a relatively low oscillation frequency with principal support inside the separation bubble, was also uncovered by the DMD analysis of the DNS results. A close up view of the low-frequency part of the spectrum for cases S20A and S40A is shown in figures \ref{fig:DMD_LowFreq}(a) and \ref{fig:DMD_LowFreq}(b), respectively.  For cases S20A and S40A (and the latter in particular) low frequency modes have relatively high amplitudes compared to other modes involved in the bubble dynamics. The eigenfunctions of the leading low frequency modes for cases S20A and S40A are shown in figures \ref{fig:low_frequency}(a) ($St = 0.18$) and \ref{fig:low_frequency}(b) ($St = 0.5$), respectively. This is particularly true for case S40A, where the $St = 0.5$ mode can also be identified as a strong peak in the $\Lambda = 40^\circ$ spectrum in figure \ref{fig:Fourier_spectra}, and could be responsible for the spanwise modulation of the 2D vortex visible in figure \ref{fig:isosurfaces}(d) near the back of the bubble. The dominant low frequency mode for case S20A has a lower frequency than for case S40A and its eigenfunction, while being three-dimensional, also shows two-dimensional features inside the bubble. The lowest frequency mode found for case S40A (not shown here) has many similarities with this mode, including a low frequency of $St = 0.12$ and a similar eigenfunction shape.

\begin{figure}
  \centerline{\includegraphics[trim = 0mm 0mm 0mm 0mm,clip,,width=1.0\textwidth]{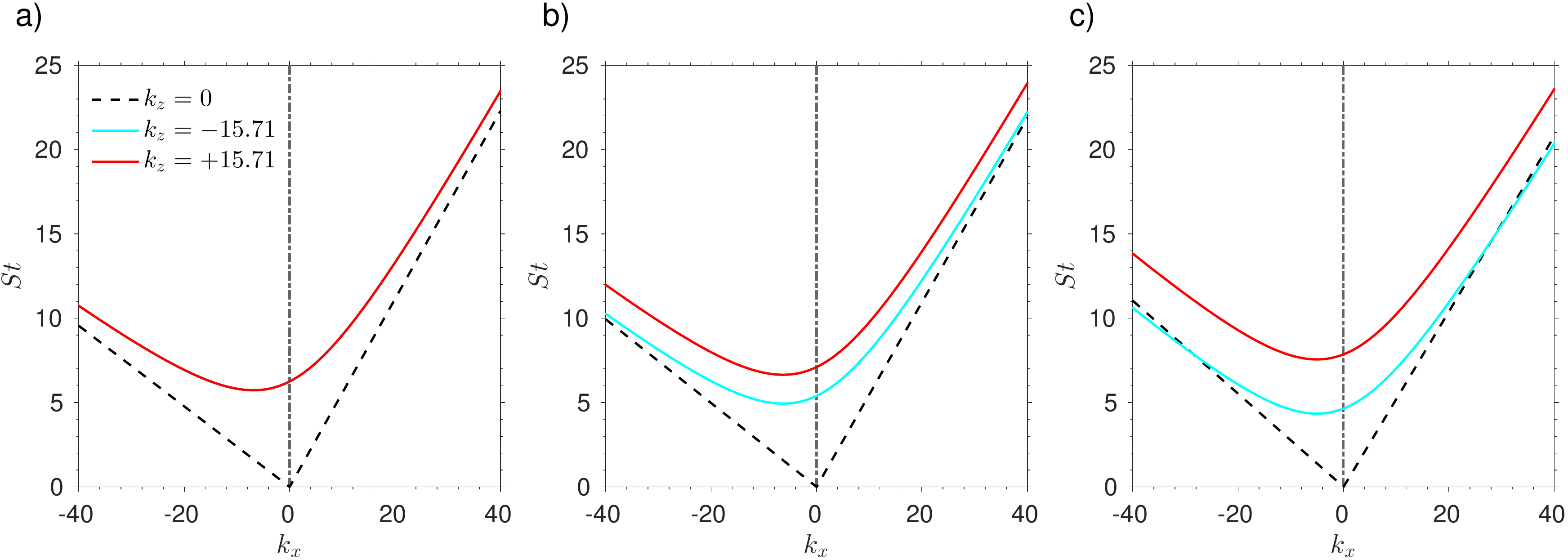}}
	\caption{Dispersion relations for oblique acoustic waves. a) case S0, b) case S20A and c) case S40A.}
\label{fig:Acoustic_relation}
\end{figure} 

Before moving on to the global analysis, we need to say something about the influence of sweep angle on the properties of upstream-propagating acoustic waves, which are known to play a role in self-excited oscillations on unswept configurations. While for the unswept case the wave vectors of the 3D modes developing in the bubble do not have a preferred orientation in the spanwise direction, the dominant 3D modes found in the swept cases always seem to have a preferred orientation. There may be different factors causing this behaviour. Instabilities will exhibit different spatial growth rates when travelling in the positive or negative spanwise direction when a nonzero mean spanwise velocity is present in the flow. Another important factor is likely to be the receptivity of the boundary layer to freestream disturbances, which can take the particular form of acoustic waves propagating upstream. When $u_z \neq 0$ relative to a generic $x-y-z$ reference frame (e.g. the airfoil reference frame), the dispersion relation for an acoustic wave is not symmetric about $k_z = 0$, hence, for a fixed frequency, the chordwise wavenumber ($k_x$) for an acoustic wave travelling backwards against the flow is different for positive and negative $k_z$. This can be shown by writing the dispersion relation for a neutral, plane acoustic wave in the free-stream in the $xyz$ reference frame (i.e. the reference frame used in the numerical simulations), which takes the form
\begin{align}
St = \displaystyle\frac{|\mathbf k|}{2\pi} \displaystyle\left[ \cos\left( \Lambda - \theta \right) +M^{-1}\right],
\label{eq:Acoustic_relation}
\end{align}
where $|\mathbf k| = \sqrt{k_x^2 + k_z^2}$, $\theta = \arctan \left( k_z/k_x \right)$ is the propagation angle in the $x-z$ plane and the sweep angle $\Lambda$ is also the angle between the $x$ direction and the free-stream flow direction. The relation (\ref{eq:Acoustic_relation}) can be easily derived by considering that the dimensionless sound speed is $c = U_\theta+1/M$ and that the flow velocity for a propagation angle $\theta$ is given by $U_\theta = U_\infty \cos\theta+W_\infty \sin\theta$. 

Figure \ref{fig:Acoustic_relation} shows the variation of $St$ against $k_x$ for $k_z = 0$ and $k_z = \pm 15.71$, which is the minimum non-zero spanwise wavenumber contained in the computational domain. While for the unswept case the dispersion relation is symmetric about $k_z$, as already anticipated the dispersion relations for $k_z> 0$ and $k_z < 0$ differ in the case of non-zero sweep. For example, at $St = 7.56$ in the $\Lambda = 40^\circ$ case, the chordwise wavenumbers for $k_z=+15.71$ are $k_x = -4.49$ and $k_x = -5.63$, while for $k_z=-15.71$ we have $k_x = -27.15$ and $k_x = 11.55$. Such a large difference in the wavenumbers of the acoustic waves travelling backwards against the flow will play a role in the receptivity process by which these disturbances are converted into instability waves. In the current case the wavenumbers of boundary layer instabilities and acoustic waves are significantly different, hence no receptivity would be expected without a wavelength conversion mechanism, for example at the leading edge or the separation point.

It can be inferred from the dispersion relations plotted in figure \ref{fig:Acoustic_relation} that modes with $k_z = \pm 15.71$ (as is the case for the low frequency modes discussed previously) and frequencies lower than about $St = 5$ cannot be sustained by free stream acoustic waves, suggesting that the low frequency modes found in the DMD are not sustained by an acoustic feedback loop. 

%


\subsection{A global stability perspective}
\begin{figure}
  \centerline{\includegraphics[trim = 0mm 0mm 0mm 0mm,clip,,width=1.0\textwidth]{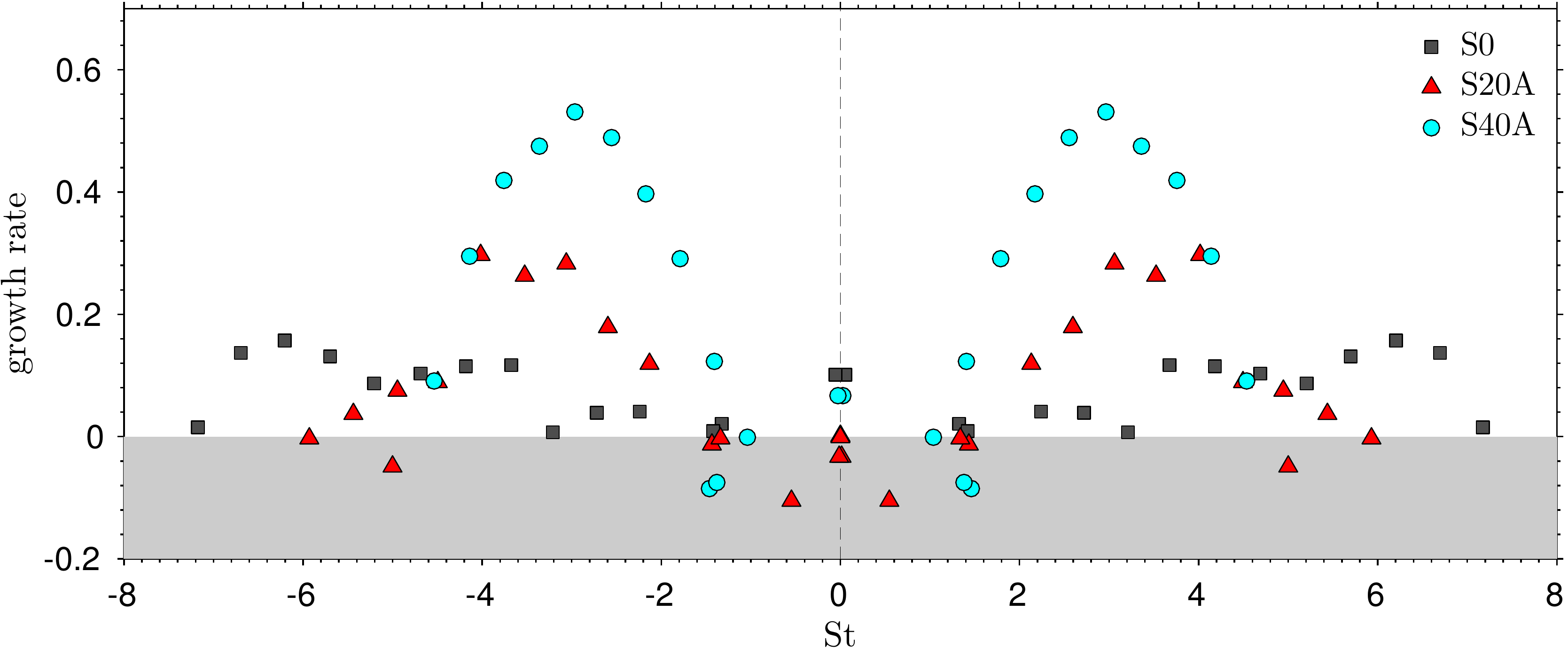}}
\caption{Global 2D spectra}
\label{global_spectra}
\end{figure}
\begin{figure}
  \centerline{\includegraphics[trim = 0mm 0mm 0mm 0mm,clip,,width=0.8\textwidth]{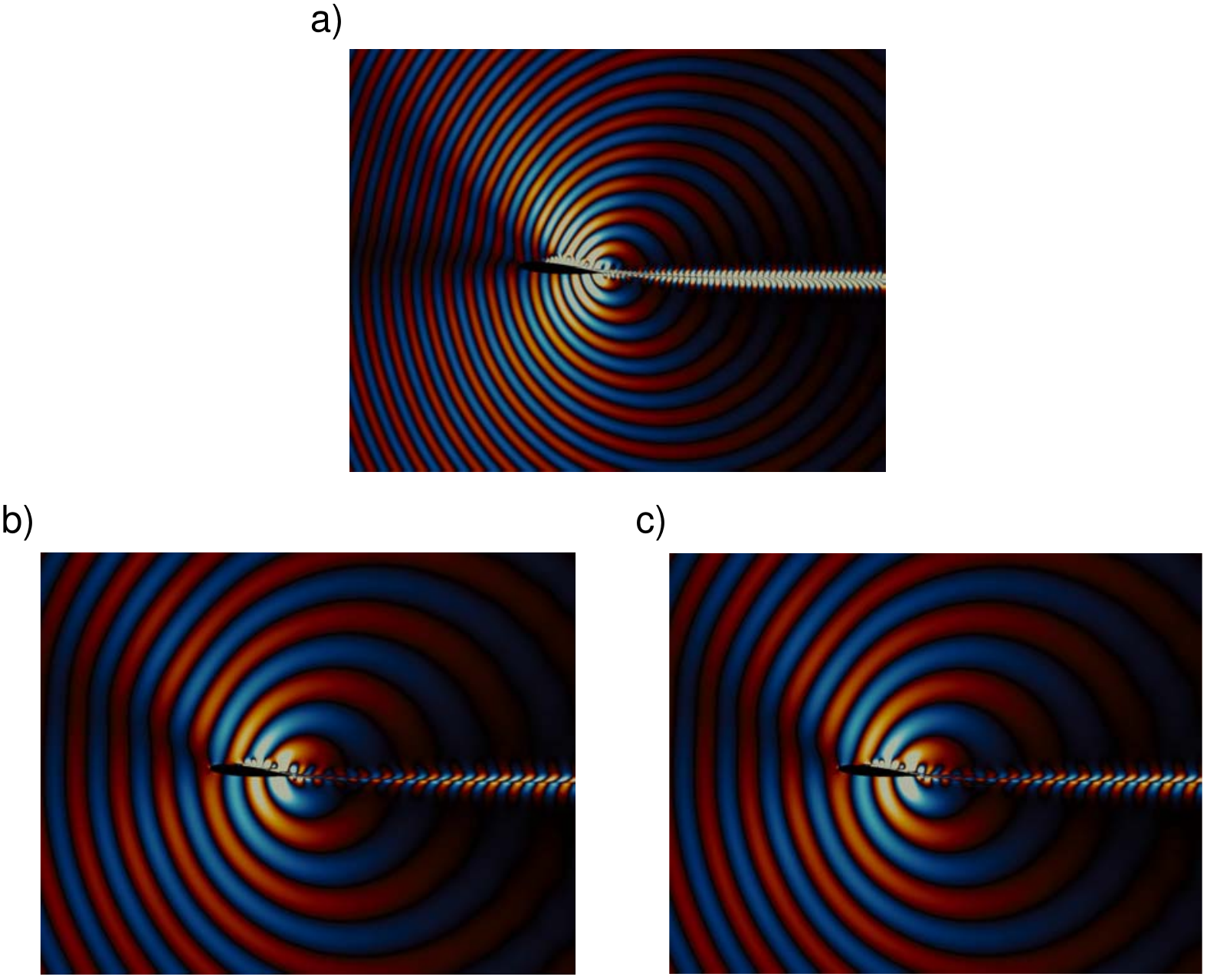}}
	\caption{Eigenfunctions of the most unstable global 2D modes shown through contours of the real part of the velocity divergence field. a) case S0 for $St = 6.20$, b) case S20A for $St = 4.0$ and c) case S40A for $St = 2.96$.}
\label{global_eigenfunctions}
\end{figure}
In addition to the dynamic mode decomposition of the DNS results, a global linear stability analysis of the time- and span-averaged flow fields was performed in order to further investigate the origin of the self-sustained vortex shedding cycle. The focus of this section is to understand the underlying mechanisms that select the dominant modes that drive the vortex shedding at high sweep angles, hence the analysis will be limited here to the two-dimensional global stability problem. Extensions to three-dimensional modes are possible, but our experience to date is that such modes are more susceptible to numerical issues in converging the global mode spectrum, whereas the two-dimensional modes shown here are fully converged.  

Figure \ref{global_spectra} shows the 2D ($k_z = 0$) global spectra for the different degrees of sweep. The mean flows obtained for all the different sweep angles are able to sustain the growth of 2D globally unstable modes. Increasing the sweep angle leads to the emergence of a strong global instability for the S40A case. A wider range of more weakly unstable modes is present for the lower sweep cases. The emergence of the strong global instability is a key result that explains the earlier observations of coherent structures and DMD modes preferentially aligned with the spanwise direction.

\begin{figure}
  \centerline{\includegraphics[trim = 0mm 0mm 0mm 0mm,clip,,width=\textwidth]{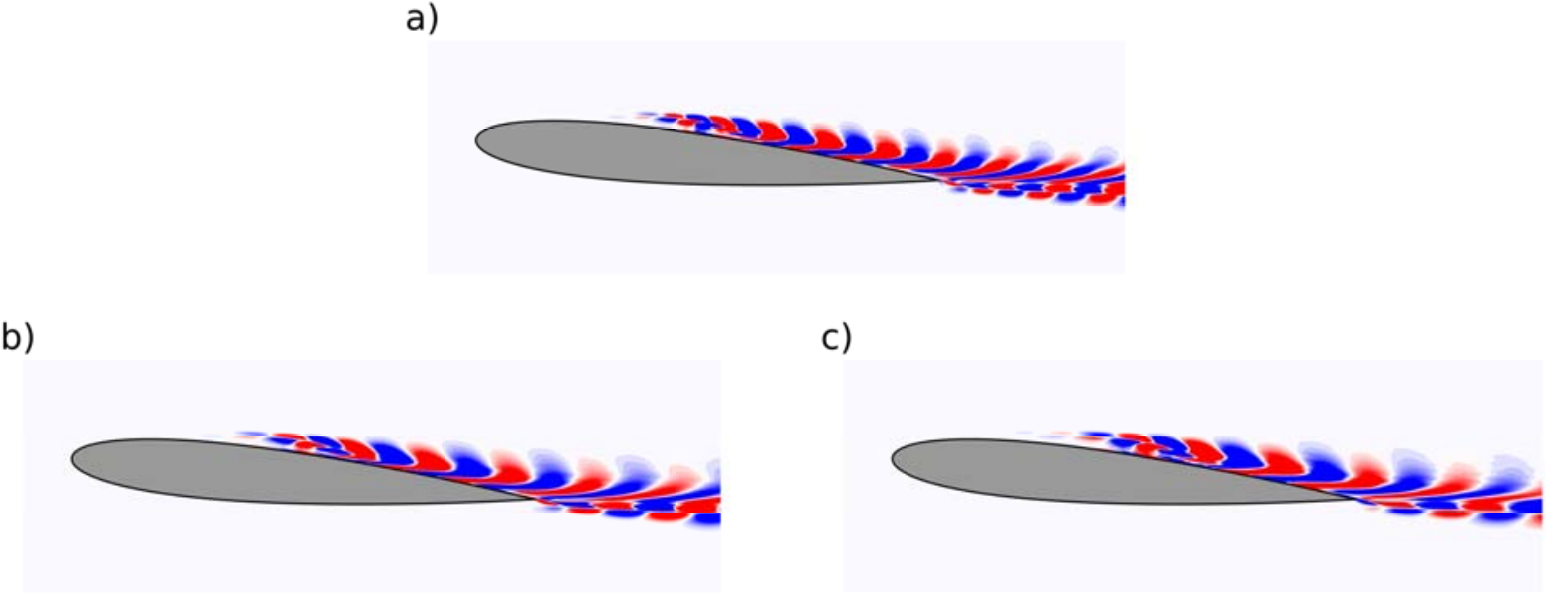}}
	\caption{Eigenfunctions of the most unstable global 2D modes shown through contours of the real part of the spanwise vorticity field. a) case S0 for $St = 6.20$, b) case S20A for $St = 4.0$ and c) case S40A for $St = 2.96$.}
\label{global_eigenfunctions_vorticity}
\end{figure}

For case S0, figure \ref{global_spectra} suggest that there are three underlying branches of instability, peaking at $St = 2.24$, $St = 3.67$ and $St = 6.20$. The second peak ($St = 3.67$) agrees well with the frequency of the strongest 2D DMD mode for this case. For case S20A, the most unstable 2D global modes occur for frequencies in the range $St = 3.0 - 4.0$. The agreement between global linear stability and DMD and Fourier analyses is good in this case, with the dominant 2D DMD and Fourier modes found for frequencies $St = 2.74$ and $St = 3.0$, respectively. For case S40A a single branch of highly unstable modes can be observed, centred at $St = 2.96$. This is in very good agreement with the DNS data; the dominant DMD and Fourier modes were found for frequencies between $St = 2.7$ and $St = 3.7$, and for $St = 3.0$, respectively. As indicated by the eigenfunctions in figure \ref{global_eigenfunctions}, showing contours of the real part of the divergence of velocity, all the most unstable global modes are characterised by a strong acoustic feedback (originating at the trailing edge). Previous global stability studies \citep{fosasdepando14} have also highlighted the importance of an acoustic feedback loop originating at the trailing edge as a source of disturbances for the excitation of shear layer modes in the separation bubble. Figure \ref{global_eigenfunctions_vorticity} shows the equivalent disturbance vorticity fields, showing that the modes also contain convective instability of the detached shear layer.
\begin{figure}
  \centerline{\includegraphics[trim = 0mm 0mm 0mm 0mm,clip,,width=\textwidth]{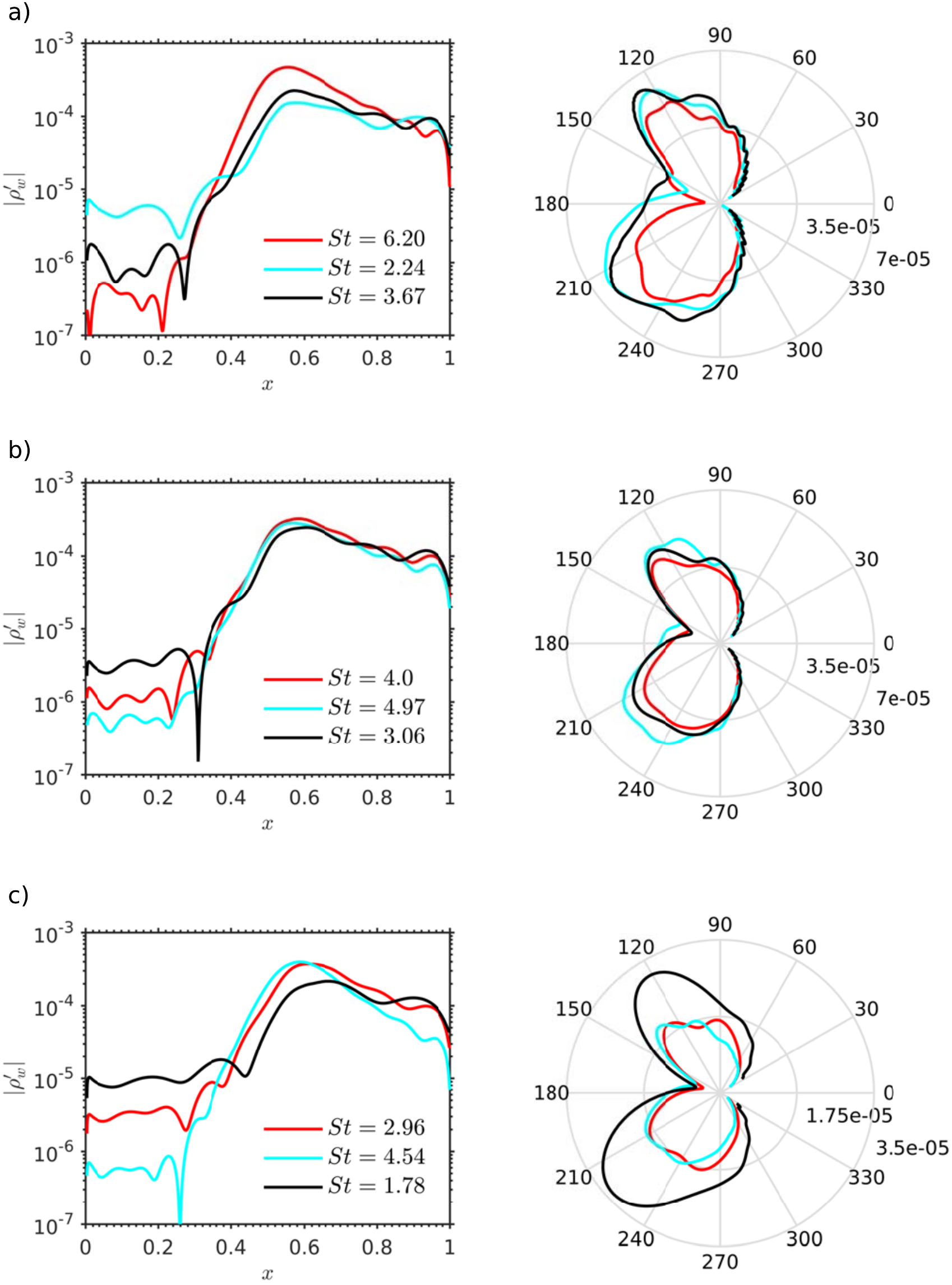}}
	\caption{Amplitude of the wall density over the suction side of the airfoil, together with polar plots of the amplitude of the divergence field in the free stream, for a selection of global modes eigenfunctions. a) case S0, b) case S20A and c) case S40A.}
\label{amplitude_polar}
\end{figure}

In order to shed some light on the mechanisms leading to the selection of the most unstable frequencies for each case, a closer inspection of the global mode eigenfunctions is provided in figure \ref{amplitude_polar}, where the amplitude of the wall density over the suction side of the airfoil is plotted, together with polar plots of the amplitude of the velocity divergence in the free stream, for a selection of Strouhal numbers. All the eigenvectors extracted from the global instability analysis are obtained in a normalised form. From the spatial structure of the global modes we can clearly identify regions of local spatial amplification of disturbances. For example, Figure \ref{amplitude_polar}(a) shows that the most unstable 2D global mode for case S0 ($St = 6.20$) is also more convectively unstable than the other dominant modes at $St = 2.24$ and $St = 3.67$, as shown by the higher gradient of the curve in the region $0.3<x<0.45$. The acoustic content is similar for the three modes, with slightly higher relative amplitudes for the divergence field at the lower frequencies. A higher leading edge receptivity coefficient for the lower frequency modes is also noticeable, indicated by the relatively larger mode amplitude ahead of the exponential growth compared to the acoustic field amplitude. The convective instability of the detached shear layer appears to dominate in terms of the overall global growth rate in this case, but the acoustic receptivity plays an important role. In fact, this combination of effects helps explain why the most globally unstable mode does not coincide with the most convectively unstable mode, which has a frequency of $St = 7.18$ and, as can be seen in figure \ref{global_spectra}, is nearly globally neutral. 

An analysis of the mode structures for case S20A is provided in figure \ref{amplitude_polar}(b). In this case, the two most unstable global modes have frequencies $St = 3.06$ and $St = 4.0$. These modes have very similar global growth rates but different local growth rates in the separated shear layer. A high receptivity coefficient appears to compensate for a low spatial growth rate for the $St = 3.06$ mode. The importance of the receptivity process is also highlighted by the behaviour of the $St = 4.97$ mode. This mode has the highest spatial growth rate of the three modes plotted but also the lowest receptivity coefficient (note that the amplitude of the divergence field in the free stream is also higher than for the other two modes), which leads to a low overall growth rate.

Figure \ref{amplitude_polar}(c) show the amplitude distribution and acoustic content of a selection of modes for case S40A. It is interesting to note that the amplitude of the acoustic feedback in the free stream (note that we are only considering acoustic waves that travel parallel to the leading edge here) decreases considerably as the sweep angle increases. The most globally unstable mode in this case has a frequency of $St = 2.96$. Again, the most globally unstable mode does not coincide with the mode showing the largest amplification across the separation bubble. As in the other cases, the selection of the most globally amplified frequencies appears to occur via a trade off between disturbance receptivity, for example near the leading edge, and strong amplitude growth in the separated shear layer.

\section{Conclusions} \label{sec:conclusions}

Direct numerical simulations were carried out to investigate the effect of sweep on the transitional separation bubbles forming on the suction side of a NACA-0012 airfoil. Two angles of sweep were investigated, in addition to the unswept case, namely $\Lambda = 20^\circ$ and $\Lambda = 40^\circ$. The simulations were all carried out for a spanwise domain size of $40\%$ chord. The analysis was carried out for two swept-wing configurations, namely a rotated wing geometry (configuration A) and a sheared geometry (configuration B). The two configurations differ in that, while configuration A maintains the standard NACA-0012 section in the leading edge perpendicular direction, this section is scaled in the chordwise direction in configuration B (see figure \ref{fig:config_sketch} for details). This leads to important differences in the flow characteristics in the two cases, due in particular to the effect of thickness. The simulations, in combination with XFoil results for the equivalent 2D airfoil sections, provide empirical support for the applicability of an independence principle.

Focusing on the results obtained for configuration A, laminar-turbulent transition of the separation bubble leads to the shedding of vortices at the back of the bubble, which is associated with a K-H instability of the detached shear layer induced by the bubble. Interestingly, even for the largest sweep angle investigated ($\Lambda = 40^\circ$) no sign of crossflow instabilities were observed. The DNS data and a two-dimensional global stability analysis of the mean flows confirms that the convective instability of the shear layer is coupled with acoustic feedback originating at the airfoil's trailing edge, consistent with previous global stability studies of flows around unswept airfoils \citep{fosasdepando14}. Here it was found that this is also true in the swept-wing case. The results of the global stability analysis were found to be in good agreement with the DNS data. In addition, the global stability data confirms that the selection of the most globally unstable frequencies is based on a trade-off between the convective instability of the separated shear layer and the acoustic receptivity near the leading edge of the airfoil.

The introduction of sweep leads to substantial modifications in the flow structure. For moderate sweep angles ($\Lambda = 20^\circ$) a single oblique mode, oriented perpendicular to the free stream dominates, while for higher sweep angles ($\Lambda = 40^\circ$) global instability results in two-dimensional (i.e.~spanwise coherent) K-H modes being dominant. An important characteristic of swept-wing flows is that, due to symmetry breaking, one orientation of oblique modes is preferred. In the moderately swept case analysed here, positive spanwise wavenumbers were found to dominate at the most amplified frequencies. This may be due to the fact that shear layer instabilities have different spatial growth rates when travelling in the positive or negative spanwise direction when there is a nonzero mean spanwise velocity in the flow. In addition to this, and perhaps more importantly, in a swept flow the chordwise wavenumber of an oblique acoustic wave travelling upstream is different, for a fixed frequency, for positive and negative spanwise wavenumbers. This, in turn, will affect the receptivity process by which disturbances enter the boundary layer. A DMD analysis carried out on the DNS results also revealed the existence of low-frequency 3D bubble modes that gain in importance as the angle of sweep is increased. These modes have principal support inside the separation bubble and are not sustained by an acoustic feedback loop. 

The present simulations have identified a number of aspects that deserve further investigation. More simulations, particularly at higher Reynolds numbers, would be needed to fully assess the change in flow structure and the relevance of the independence principle for swept flows, whereas simulations with wider spanwise domains are needed to explore the presence of global modes with even longer spanwise wavelengths than those identified in the present simulations. The influence on the global modes of the inhomogeneous flow near wing tips also remains to be studied.

\begin{center}
\section*{Acknowledgments}
\end{center}
The authors would like to acknowledge support from EPSRC under grants EP/M822692/1 and EP/L000261/1.

\bibliographystyle{jfm}
\bibliography{my_bib}

\end{document}